\documentclass[aps,prd,twocolumn,amssymb,eqsecnum,showpacs,nofootinbib]{revtex4}
\usepackage{epsfig}
\usepackage{amsmath}
\usepackage{amssymb}

\begin{document}

\begin{widetext}

For reference, the following erratum corrects the published version of the paper. These errors have been fixed in this arxiv-version (the article starting on page 2 has the corrected expressions).
 
\medskip

\begin{center}
{\large \bf{
Erratum: Evolution of the Carter constant for inspirals into a black hole: Effect of the black hole quadrupole\\
\ [Phys.\ Rev.\ D {\bf 75}, 124007 (2007)]}
}

\medskip

\'Eanna \'E.\ Flanagan, Tanja Hinderer
\end{center}

\medskip

In Eqs.\ (3.16), (3.17), (3.18), (3.24), (3.25) and (3.26) of this paper, the variable $r$ should be replaced everywhere by the variable ${\tilde r}$, and the variable $\theta$ should be replaced everywhere by the variable ${\tilde \theta}$.  The definitions of ${\tilde r}$ and ${\tilde \theta}$ are given in Eq.\ (2.11).  These replacements do not affect the any of the subsequent results in the paper.

Also, the right hand side of Eq.\ (B3) is missing a term $- 4 S L_z
\tilde r$ and Eq. (2.24) is missing a factor of $d\varphi/d\tilde t$ in front of $Q$.

Some terms are missing in Eqs.\ (3.18), (3.26) and (3.30) - (3.33).
The additional terms in Eqs.\ (3.18) and (3.26) are
\begin{equation}
-\frac{8 Q}{15 \tilde r^7} \left[-75 K^2+2
  K \tilde r (51 \tilde r E+50)+8 \tilde r^2 (\tilde r E+1) (3 \tilde
  r E+5)\right],
\nonumber
\end{equation}
and 
\begin{equation}
\frac{8 Q}{15 p^2 \tilde r^7} \left[25 p^3
  (3 p - 4\tilde r)+p^2 \tilde r^2 \left(11-51 e^2\right)+32 p \tilde
  r^3 \left(1-e^2\right)+6 \tilde r^4 \left(1-e^2\right)^2\right],\nonumber
\end{equation}
respectively. These result in additional fractional corrections to
Eq. (3.30) given by
\begin{equation}
-\frac{Q}{p^2}\left[\frac{1}{2}+\frac{73}{48}e^2+\frac{37}{192}e^4\right],\nonumber
\end{equation}
and the full expression replacing the $O(Q)$ terms in Eq. (3.30) is then
\begin{equation}
\langle \dot K\rangle=-\frac{64}{5}\frac{(1-e^2)^{3/2}}{p^3}\left[1+\frac{7 e^2}{8}-\frac{Q }{p^2}\left\{1+\frac{8}{3}e^2+\frac{11}{12}e^4+\left(\frac{13}{4}+\frac{841}{96}e^2+\frac{449}{192}e^4\right) \cos (2
   \iota )\right\}\right]+O(S),O(S^2)-{\rm terms}.\nonumber
\end{equation}
Equations (3.31), (3.32) and (3.33) contain typos in the $O(S)$ and $O(Q)$ terms, the corrected expressions are given below. We thank P. Komorowski for pointing this out.  Equation (3.31) should be replaced by
\begin{eqnarray}
 \langle \dot p\rangle &=& -\frac{64}{5}\frac{(1-e^2)^{3/2}}{p^3}\left\{1+\frac{7e^2}{8}-\frac{S\cos(\iota)}{96p^{3/2}}\left(1064+1516e^2+475e^4\right)\right.\nonumber\\
&& \ \ \ \ \ \ \ \ \ \ \ \ -\frac{Q}{8p^{2}}\left[14+\frac{149e^2}{12}+\frac{19e^4}{48}+\left(50+\frac{469e^2}{12}+\frac{227 e^4}{24}\right)\cos(2\iota)\right]\nonumber\\
&&\ \ \ \ \ \ \ \   \ \ \ +\left.\frac{S^2}{64 p^2}\left(\frac{1}{3}+e^2+\frac{e^4}{8}\right)\left[13-\cos(2\iota)\right]\right\}, \ \ \ \ \ \ \ \ \ \  \ \ \ \ \ \ \ \ \ \ \ \ \ 
\end{eqnarray}
Equation (3.32) should be replaced by
\begin{eqnarray}
 \langle\dot e\rangle&=& -\frac{304}{15}\frac{e(1-e^2)^{3/2}}{p^4}\left(1+\frac{121e^2}{304}\right)+\frac{Se(1-e^2)^{3/2}\cos(\iota)}{5 p^{11/2}}\left(1172+932e^2+\frac{1313e^4}{6}\right)\nonumber\\
&& +\frac{Q(1-e^2)^{3/2}}{ ep^6}\left[32+\frac{785 e^2}{3}-\frac{219e^4}{2}+13e^6+\left(32+\frac{2195e^2}{3}+251e^4+\frac{218e^6}{3}\right)\cos(2\iota)\right]\nonumber\\
&&-\frac{S^2e(1-e^2)^{3/2}}{8 p^6}\left(2+3e^2+\frac{e^4}{4}\right)\left[13-\cos(2 \iota)\right],
\end{eqnarray}
and the corrected Eq. (3.33) is
\begin{eqnarray}
  \langle \dot \iota\rangle &=& \frac{S\sin (\iota) (1-e^2)^{3/2}}{ p^{11/2}}\left[\frac{244}{15}+\frac{252}{5}e^2+\frac{19}{2}e^4\right]-\frac{\left(1-e^2\right)^{3/2}  S^2  \sin (2 \iota )}{240
   p^6}\left[8+3 e^2\left(8+e^2\right)\right]\nonumber\\
&&+\frac{Q\cot(\iota)(1-e^2)^{3/2}}{60p^{6}}\left[312+736 e^2-83 e^4-\left(408+1268 e^2+599e^4\right)\cos(2\iota)\right].
\end{eqnarray}

\newpage

\end{widetext}

\bigskip

\title{Evolution of the Carter constant for inspirals into a black hole: effect of the black hole quadrupole}

\date{\today}
\author{ \'Eanna \'E. Flanagan$^{1,2}$}
\author{Tanja Hinderer$^1$} \affiliation{$^1$ Center for Radiophysics and Space
Research, Cornell University, Ithaca, NY 14853, USA\\
$^2$ Laboratory for
Elementary Particle Physics, Cornell University, Ithaca, NY 14853,
USA}
\def\be{\begin{equation}}
\def\ee{\end{equation}}
\def\bea{\begin{eqnarray}}
\def\eea{\end{eqnarray}}
\newcommand{\bes}{\begin{subequations}}
\newcommand{\ees}{\end{subequations}}
\begin{abstract}
We analyze the effect of gravitational radiation reaction on
generic orbits around a body with an axisymmetric mass quadrupole
moment $Q$ to linear order in $Q$, to the leading post-Newtonian
order, and to linear order in the mass ratio. This system admits
three constants of the motion in absence of radiation reaction:
energy, angular momentum along the symmetry axis, and a third
constant analogous to the Carter constant. We compute
instantaneous and time-averaged rates of change of these three
constants. For a point particle orbiting a black hole, Ryan
\cite{ryan2} has computed the leading order evolution of the
orbit's Carter constant, which is linear in the spin. Our result,
when combined with an interaction quadratic in the spin (the
coupling of the black hole's spin to its own radiation reaction
field), gives the next to leading order evolution. The effect of
the quadrupole, like that of the linear spin term, is to
circularize eccentric orbits and to drive the orbital plane
towards antialignment with the symmetry axis.

In addition we consider a system of two point masses where one body
has a single mass multipole or current multipole of order $l$. To
linear order in the mass ratio, to linear order in the multipole,
and to the leading post-Newtonian order, we show that there does not
exist an analog of the Carter constant for such a system (except for
the cases of an $l=1$ current moment and an $l=2$ mass moment).
Thus, the existence of the Carter constant in Kerr depends on
interaction effects between the different multipoles.
With mild additional assumptions, this result falsifies the conjecture
that all vacuum, axisymmetric spacetimes posess a third constant of the motion
for geodesic motion.

\end{abstract}

\maketitle

\section{introduction and summary}
\label{intro} The inspiral of stellar mass compact objects with
masses $\mu$ in the range $\mu\sim 1-100M_\odot$ into massive black
holes with masses $M\sim 10^5-10^7M_\odot$ is one of the most
important sources for the future space-based gravitational wave
detector LISA. Observing such events will provide a variety of
information: (i) the masses and spins of black holes can be measured
to high accuracy $(\sim 10^{-4})$; which can constrain the black
hole's growth history \cite{sources}; (ii) the observations will
give a precise test of general relativity in the strong field regime
and unambiguously identify whether the central object is a black
hole \cite{mapgeo}; and (iii) the measured event rate will give
insight into the complex stellar dynamics in galactic nuclei
\cite{sources}. Analogous inspirals may also be interesting for the
advanced stages of ground-based detectors: it has been estimated
that advanced LIGO could detect up to $\sim 10-30$ inspirals per
year of stellar mass compact objects into intermediate mass black
holes with masses $M\sim 10^2-10^4M_\odot$ in globular clusters
\cite{imris}. Detecting these inspirals and extracting information
from the datastream will require accurate models of the
gravitational waveform as templates for matched filtering. For
computing templates, we therefore need a detailed understanding of
the how radiation reaction influences the evolution of bound orbits
around Kerr black holes \cite{Glampedakis,lrr,cqg,steve}.

There are three dimensionless parameters characterizing inspirals of
bodies into black holes:
\begin{itemize}
\item the dimensionless spin parameter $a=|{\bf{S}}|/M^2$ of the black hole, where ${\bf S}$ is the spin.
\item the strength of the interaction potential $\epsilon^2=
GM/rc^2$, i.e.\ the expansion parameter used in post-Newtonian (PN)
theory.
\item the mass ratio $\mu/M$.
\end{itemize}
For LISA data analysis we will need waveforms that are accurate to
all orders in $a$ and $\epsilon^2$, and to leading order in $\mu/M$.
However, it is useful to have analytic results in the regimes $a \ll
1$ and/or $\epsilon^2 \ll 1$.  Such approximate results can be
useful as a check of numerical schemes that compute more accurate
waveforms, for scoping out LISA's data analysis requirements
\cite{Glampedakis,sources}, and for assessing the accuracy of the
leading order in $\mu/M$ or adiabatic approximation
\cite{scalar,Marc,res1}. There is substantial literature on such
approximate analytic results, and in this paper we will extend some
of these results to higher order.

A long standing difficulty in computing the evolution of generic
orbits has been the evolution of the orbit's "Carter constant", a
constant of motion which governs the orbital shape and inclination.
A theoretical prescription now exists for computing Carter constant
evolution to all orders in $\epsilon$ and $a$ in the adiabatic limit
$\mu \ll M$ \cite{Mino2003,Qdot,PRL,scalar}, but it has not yet been
implemented numerically. In this paper we focus on computing
analytically the evolution of the Carter constant in the regime $a
\ll 1$, $\epsilon \ll 1$, $\mu/M \ll 1$, extending earlier results
by Ryan \cite{ryan1,ryan2}.

We next review existing analytical work on the effects of multipole
moments on inspiral waveforms. For non-spinning point masses, the
phase of the $l=2$ piece of the waveform is known to $O(\epsilon^7)$
beyond leading order \cite{pn72}, while spin corrections are not
known to such high order. To study the leading order effects of the
central body's multipole moments on the inspiral waveform, in the
test mass limit $\mu \ll M$, one has to correct both the
conservative and dissipative pieces of the forces on the bodies. For
the conservative pieces, it suffices to use the Newtonian action for
a binary with an additional multipole interaction potential. For the
dissipative pieces, the multipole corrections to the fluxes at
infinity of the conserved quantities can simply be added to the
known PN point mass results. The lowest order spin-orbit coupling
effects on the gravitational radiation were first derived by Kidder
\cite{Kidder}, then extended by Ryan \cite{ryan1,ryan2}, Gergely
\cite{gergelyspin}, and Will \cite{will}. Recently, the corrections
of $O(\epsilon^2)$ beyond the leading order to the spin-orbit
effects on the fluxes were derived \cite{spineom,spinrad}.
Corrections to the waveform due to the quadrupole - mass monopole
interaction were first considered by Poisson \cite{poisson}, who
derived the effect on the time averaged energy flux for circular
equatorial orbits. Gergely \cite{Gergely} extended this work to
generic orbits and computed the radiative instantaneous and time
averaged rates of change of energy $E$, magnitude of angular
momentum $|{\bf L}|$, and the angle $\kappa=\cos^{-1}({\bf S}\cdot
{\bf L})$ between the spin ${\bf S}$ and orbital angular momentum
${\bf L}$. Instead of the Carter constant, Gergely identified the
angular average of the magnitude of the orbital angular momentum,
$\bar L$, as a constant of motion. The fact that to post-2-Newtonian
(2PN) order there is no time averaged secular evolution of the spin
allowed Gergely to obtain expressions for $\dot L$ and $\dot \kappa$
from the quadrupole formula for the evolution of the total angular
momentum ${\bf J}={\bf L}+{\bf S}$. In a different paper, Gergely
\cite{gergelyspin} showed that in addition to the quadrupole,
self-interaction spin effects also contribute at $2$PN order, which
was seen previously in the black hole perturbation calculations of
Shibata et al. \cite{pnkerr}. Gergely calculated the effect of this
interaction on the instantaneous and time-averaged fluxes of $E$ and
$|{\bf L}|$ but did not derive the evolution of the third constant
of motion.

In this paper, we will re-examine the effects of the quadrupole
moment of the black hole and of the leading order spin self
interaction. For a black hole, our analysis will thus contain all
effects that are quadratic in spin to the leading order in
$\epsilon^2$ and in $\mu/M$. Our work will extend earlier work by
\begin{itemize}
\item Considering generic orbits. \item Using a natural
generalization of the Carter-type constant that can be defined for
two point particles when one of them has a quadrupole. This
facilitates applying our analysis to Kerr inspirals. \item
Computing instantaneous as well as time-averaged fluxes for all
three constants of motion: energy $E$, $z$-component of angular
momentum $ L_z$, and Carter-type constant $ K$. For most purposes,
only time-averaged fluxes are needed as only they are gauge
invariant and physically relevant. However, there is one effect
for which the time-averaged fluxes are insufficient, namely
transient resonances that occur during an inspiral in Kerr in the
vicinity of geodesics for which the radial and azimuthal
frequencies are commensurate \cite{res,res1}. The instantaneous
fluxes derived in this paper will be used in \cite{res1} for
studying the effect of these resonances on the gravitational wave
phasing.
\end{itemize}

We will analyze the effect of gravitational radiation reaction on
orbits around a body with an axisymmetric mass quadrupole moment $Q$
to leading order in $Q$, to the leading post-Newtonian order, and to
leading order in the mass ratio. With these approximations the
adiabatic approximation holds: gravitational radiation reaction
takes place over a timescale much longer than the orbital period, so
the orbit looks geodesic on short timescales. We follow Ryan's
method of computation \cite{ryan1}: First, we calculate the orbital
motion in the absence of radiation reaction and the associated
constants of motion. Next, we use the leading order radiation
reaction accelerations that act on the particle (given by the
Burke-Thorne formula \cite{MTW} augmented by the relevant spin
corrections \cite{ryan1}) to compute the evolution of the constants
of motion. In the adiabatic limit, the time-averaged rates of change
of the constants of motion can be used to infer the secular orbital
evolution. Our results show that a mass quadrupole has the same
qualitative effect on the evolution as spin: it tends to circularize
eccentric orbits and drive the orbital plane towards antialignment
with the symmetry axis of the quadrupole.

The relevance of our result to point particles inspiralling into
black holes is as follows. The vacuum spacetime geometry around any
stationary body is completely characterized by the body's mass
multipole moments $I_L=I_{a_1, a_2 \ldots a_l}$ and current
multipole moments $S_L=S_{a_1, a_2 \ldots a_l}$ \cite{Hansen}. These
moments are defined as coefficients in a power series expansion of
the metric in the body's local asymptotic rest frame {\cite{RMP}}.
For nearly Newtonian sources, they are given by integrals over the
source as \bea I_L&\equiv& I_{a_1, \ldots a_l}=\int  \rho
x_{<a_1}\ldots x_{a_l>}d^3 x,\\
 S_L&\equiv& S_{a_1, \ldots a_l}=\int
\rho x_p v_{q} \epsilon_{ pq<a_1}x_{a_2}\ldots x_{a_l>}d^3 x . \eea
Here $\rho$ is the mass density and $v_q$ is the velocity, and "$<
\dots>$" means "symmetrize and remove all traces". For axisymmetric
situations, the tensor multipole moments $I_L$ ($S_L$) contain only
a single independent component, conventionally denoted by $I_l$
($S_l$) \cite{Hansen}. For a Kerr black hole of mass $M$ and spin
${\bf S}$, these moments are given by \cite{Hansen} \be I_l+ iS_l =
M^{l+1}(ia)^l, \label{bhmulti}\ee where $a$ is the dimensionless
spin parameter defined by $a=|{\bf S}|/M^2$. Note that $S_l=0$ for
even $l$ and $I_l=0$ for odd $l$.

Consider now inspirals into an axisymmetric body which has some
arbitrary mass and current multipoles $I_l$ and $S_l$. Then we can
consider effects that are linear in $I_l$ and $S_l$ for each $l$,
effects that are quadratic in the multipoles proportional to
$I_lI_{l'}$, $I_lS_{l'}$, $S_lS_{l'}$, effects that are cubic, etc.
For a general body, all these effects can be separated using their
scalings, but for a black hole, $I_l\propto a^l$ for even $l$ and
$S_l\propto a^l$ for odd $l$ [see Eq.(\ref{bhmulti})], so the
effects cannot be separated. For example, a physical effect that
scales as $O(a^2)$ could be an effect that is quadratic in the spin
or linear in the quadrupole; an analysis in Kerr cannot distinguish
these two possibilities. For this reason, it is useful to analyze
spacetimes that are more general than Kerr, characterized by
arbitrary $I_l$ and $S_l$, as we do in this paper. For recent work
on computing exact metrics characterized by sets of moments $I_l$
and $S_l$, see Refs. \cite{spacetime1, spacetime2} and references
therein.

The leading order effect of the black hole's multipoles on the
inspiral is the $O(a)$ effect computed by Ryan \cite{ryan2}. This
$O(a)$ effect depends linearly on the spin $S_1$ and is independent
of the higher multipoles $S_l$ and $I_l$ since these all scale as
$O(a^2)$ or smaller. In this paper we compute the $O(a^2)$ effect on
the inspiral, which includes the leading order linear effect of the
black hole's quadrupole (linear in $I_2\equiv Q$) and the leading
order spin self-interaction (quadratic in $S_1$).

We next discuss how these $O(a^2)$ effects scale with the
post-Newtonian expansion parameter $\epsilon$. Consider first the
conservative orbital dynamics. Here it is easy to see that
fractional corrections that are linear in $I_2$ scale as
$O(a^2\epsilon^4)$, while those quadratic in $S_1$ scale as
$O(a^2\epsilon^6)$. Thus, the two types of terms cleanly separate.
We compute only the leading order, $O(a^2\epsilon^4)$, term. For the
dissipative contributions to the orbital motion, however, the
scalings are different. There are corrections to the radiation
reaction acceleration whose fractional magnitudes are
$O(a^2\epsilon^4)$ from both types of effects linear in $I_2$ and
quadratic in $S_1$. The effects quadratic in $S_1$ are due to the
backscattering of the radiation off the piece of spacetime curvature
due to the black hole's spin. This effect was first pointed out by
Shibata et al. {\cite{pnkerr}}, who computed the time-averaged
energy flux for circular orbits and small inclination angles based
on a PN expansion of black hole perturbations. Later, Gergely
{\cite{gergelyspin}} analyzed this effect on the instantaneous and
time-averaged fluxes of energy and magnitude of orbital angular
momentum within the PN framework.

The organization of this paper is as follows. In Sec. \ref{sec1},
we study the conservative orbital dynamics of two point particles
when one particle is endowed with an axisymmetric quadrupole, in
the weak field regime, and to leading order in the mass ratio. In
Sec. \ref{sec2}, we compute the radiation reaction accelerations
and the instantaneous and time-averaged fluxes. In order to have
all the contributions at $O(a^2\epsilon^4)$ for a black hole, we
include in our computations of radiation reaction acceleration the
interaction that is quadratic in the spin $S_1$. The application
to black holes in Sec. \ref{sec3} briefly discusses the
qualitative predictions of our results and also compares with
previous results.

The methods used in this paper can be applied only to the black hole
spin (as analyzed by Ryan \cite{ryan1}) and the black hole
quadrupole (as analyzed here). We show in Sec. \ref{sec4} that for
the higher order mass and current multipole moments taken
individually, an analog of the Carter constant cannot be defined to
the order of our approximations.  We
then show that under mild assumptions, this non-existence result
can be extended to exact spacetimes, thus falsifying the conjecture
that all vacuum axisymmetric spacetimes possess a third constant of
geodesic motion.

\section{effect of an axisymmetric mass quadrupole on the conservative orbital dynamics}
\label{sec1}

Consider two point particles $m_1$ and $m_2$ interacting in
Newtonian gravity, where $m_2 \ll m_1$ and where the mass $m_1$ has
a quadrupole moment $Q_{ij}$ which is axisymmetric: \bea Q_{ij}& =&
\int d^3 x \rho({\bf r}) \left[ x_i x_j - \frac{1}{3} r^2
  \delta_{ij} \right] \\
  &= &Q \left( n_i n_j - \frac{1}{3} \delta_{ij}\right).
\eea  For a Kerr black hole of mass $M$ and dimensionless spin
parameter $a$ with spin axis along ${\bf n}$, the quadrupole scalar
is $Q=-M^3a^2$.

The action describing this system, to leading order in $m_2/m_1$, is
\be S = \int dt \left[ \frac{1}{2}\mu {\bf v}^2 - \mu \Phi({\bf r})
\right], \label{action} \ee where ${\bf v} = {\dot {\bf r}}$ is the
velocity, the potential is \be \Phi({\bf r}) = - \frac{M}{r} -
\frac{3}{2 r^5} x^i x^j Q_{ij}, \ee $\mu$ is the reduced mass and
$M$ the total mass of the binary, and we are using units with $G
=c=1$. We work to linear order in $Q$, to linear order in $m_2/m_1$,
and to leading order in $M/r$. In this regime, the action
(\ref{action}) also describes the conservative effect of the black
hole's mass quadrupole on bound test particles in Kerr, as discussed
in the introduction. We shall assume that the quadrupole $Q_{ij}$ is
constant in time. In reality, the quadrupole will evolve due to
torques that act to change the orientation of the central body. An
estimate based on treating $m_1$ as a rigid body in the Newtonian
field of $m_2$ gives the scaling of the timescale for the quadrupole
to evolve compared to the radiation reaction time as (see Appendix I
for details) \be \frac{T_{\rm evol}}{T_{\rm
rr}}\sim\left(\frac{m_1}{m_2}\right)\left(\frac{M}{r}\right)\left(\frac{\bar
S}{\bar Q}\right)\sim \left(\frac{M}{\mu}\right)
\left(\frac{M}{r}\right)\left(\frac{1}{a}\right).\label{tevol}\ee
Here, we have denoted the dimensionless spin and quadrupole of the
body by $\bar S$ and $\bar Q$ respectively, and the last relation
applies for a Kerr black hole. Since $\mu/M\ll 1$, the first factor
in Eq. (\ref{tevol}) will be large, and since $1/a\geq 1$ and for
the relativistic regime $M/r\sim 1$, the evolution time is long
compared to the radiation reaction time. Therefore we can neglect
the evolution of the quadrupole at leading order.

This system admits three conserved quantities, the energy \be E =
\frac{1}{2} \mu {\bf v}^2 + \mu \Phi({\bf r}),\label{ecart} \ee the
$z$-component of angular momentum \be L_z = {\bf e}_z \cdot (\mu
{\bf r} \times {\bf v}), \ee and the Carter-type constant \bea K &=&
\mu^2 ( {\bf r} \times {\bf v})^2 - \frac{2 Q \mu^2}{r^3} ( {\bf n}
\cdot {\bf r} )^2 \nonumber\\
&&+ \frac{Q \mu^2}{M} \left[ ({\bf n}\cdot{\bf v})^2 - \frac{1}{2}
{\bf v}^2 + \frac{M}{r} \right]. \label{kcartesianq}\eea (See below
for a derivation of this expression for $K$).

\subsection{Conservative orbital dynamics in a Boyer-Lindquist-like coordinate system}

We next specialize to units where $M=1$.  We also define the
rescaled conserved quantities by $\tilde E = E/\mu$, $\tilde L_z =
L_z/\mu$, $\tilde K = K / \mu^2$, and drop the tildes. These
specializations and definitions have the effect of eliminating all
factors of $\mu$ and $M$ from the analysis. In spherical polar
coordinates $(r,\theta,\varphi)$ the constants of motion $E$ and
$L_z$ become
\begin{eqnarray}
E&=&\frac{1 }{ 2} ({\dot r}^2 + r^2 {\dot \theta}^2 + r^2
\sin^2\theta {\dot \varphi}^2) - \frac{1}{r} \nonumber\\
&&+ \frac{Q
}{2 r^3} (1 - 3 \cos^2 \theta), \\
L_z &=& r^2 \sin^2 \theta {\dot \varphi}.
\end{eqnarray}
In these coordinates, the Hamilton-Jacobi equation is not
separable, so a separation constant $K$ cannot readily be derived.
For this reason we switch to a different coordinate system
$({\tilde r}, {\tilde \theta}, \varphi)$ defined by \bea r
\cos\theta &=& {\tilde r} \cos{\tilde \theta} \left( 1 +
\frac{Q}{4
    {\tilde r}^2} \right), \nonumber \\
r \sin\theta &=& {\tilde r} \sin{\tilde \theta} \left( 1 - \frac{Q}{4
    {\tilde r}^2} \right).
\label{tildecoord}\eea We also define a new time variable ${\tilde
t}$ by \be dt = \left[ 1 - \frac{Q}{2 {\tilde r}^2} \cos(2 {\tilde
\theta}) \right] d{\tilde t}.\label{ttildedef} \ee

The action (\ref{action}) in terms of the new variables to linear
order in $Q$ is \bea S &=&\int d{\tilde t}\left\{ \frac{1}{2} \left(
\frac{d {\tilde r}}{d{\tilde t}}\right)^2 + \frac{1}{2} {\tilde r}^2
\left( \frac{d {\tilde \theta}}{d{\tilde
      t}}\right)^2\right.\nonumber\\
&&{\;}{\;}{\;}{\;}{\;}{\;}{\;}{\;}{\;}+ \left.\frac{1}{2} {\tilde
r}^2 \sin^2 {\tilde \theta} \left( \frac{d \varphi}{d{\tilde
      t}}\right)^2 \left[ 1 - \frac{Q}{{\tilde r}^2} \sin^2 {\tilde
    \theta} \right] \right.\nonumber\\
    &&{\;}{\;}{\;}{\;}{\;}{\;}{\;}{\;}{\;}+ \left.\frac{1}{{\tilde r}} + \frac{Q}{4 {\tilde r}^3} \right\}.
\label{action1} \eea However, a difficulty is that the action
(\ref{action1}) does not give the same dynamics as the original
action (\ref{action}).  The reason is that for solutions of the
equations of motion for the action (\ref{action}), the variation of
the action vanishes for paths with fixed endpoints for which the
time interval $\Delta t$ is fixed. Similarly, for solutions of the
equations of motion for the action (\ref{action1}), the variation of
the action vanishes for paths with fixed endpoints for which the
time interval $\Delta {\tilde t}$ is fixed. The two sets of varied
paths are not the same, since $\Delta t \ne \Delta {\tilde t}$ in
general. Therefore, solutions of the Euler-Lagrange equations for
the action (\ref{action}) do not correspond to solutions of the
Euler-Lagrange equations for the action (\ref{action1}). However, in
the special case of zero-energy motions, the extra terms in the
variation of the action vanish. Thus, a way around this difficulty
is to modify the original action to be \be \hat S = \int dt \left[
\frac{1}{2}\mu {\bf v}^2 - \mu \Phi({\bf r}) +E\right].
\label{action2} \ee This action has the same extrema as the action
(\ref{action}), and for motion with physical energy $E$, the energy
computed with this action is zero. Transforming to the new variables
yields, to linear order in $Q$: \bea \hat S &=& \int d{\tilde t}
\left\{ \frac{1}{2} \left( \frac{d {\tilde r}}{d{\tilde t}}\right)^2
+ \frac{1}{2} {\tilde r}^2 \left( \frac{d {\tilde \theta}}{d{\tilde
      t}}\right)^2\right.\nonumber\\
&&{\;}{\;}{\;}{\;}{\;}{\;}{\;}{\;}{\;}+ \left.\frac{1}{2} {\tilde
r}^2 \sin^2 {\tilde \theta} \left( \frac{d \varphi}{d{\tilde
      t}}\right)^2 \left[ 1 - \frac{Q}{{\tilde r}^2} \sin^2 {\tilde
    \theta} \right]\right.\nonumber\\
    &&{\;}{\;}{\;}{\;}{\;}{\;}{\;}{\;}{\;}+\left.\frac{1}{{\tilde r}} +\frac{Q}{4 {\tilde r}^3}
+ E - \frac{Q E}{2 {\tilde r}^2} \cos(2 {\tilde \theta})\right\}.
\label{action3} \eea The zero-energy motions for this action
coincide with the zero energy motions for the action
(\ref{action2}). We use this action (\ref{action3}) as the
foundation for the remainder of our analysis in this section.

The $z$-component of angular momentum in terms of the new variables
$(\tilde r, \tilde \theta, \varphi, \tilde t)$ is \be L_z = {\tilde
r}^2 \sin^2 {\tilde \theta} \left( \frac{d \varphi}{d{\tilde
      t}}\right) \left[ 1 - \frac{Q}{{\tilde r}^2} \sin^2 {\tilde
    \theta} \right].
\label{s3q} \ee  We now transform to the Hamiltonian: \bea \hat
H&=&\frac{1}{2}p^2_{\tilde r}-\frac{1}{\tilde r}-E-\frac{Q}{4\tilde
r^3}+\frac{QL_z^2}{2\tilde r^4}\nonumber\\
&&+\frac{1}{2\tilde r^2}\left[p^2_{\tilde
\theta}+\frac{L^2_z}{\sin^2\tilde \theta}+QE\cos (2 \tilde
\theta)\right]\label{hhat} \eea and solve the Hamiltonian Jacobi
equation. Denoting the separation constant by $K$ we obtain the
following two equations for the $\tilde r$ and $\tilde \theta$
motions: \be \label{s1q} \left( \frac{d \tilde r}{d \tilde t}
\right)^2 = 2 E + \frac{2}{\tilde
  r} - \frac{K}{{\tilde r}^2} + \frac{Q}{2} \left[ \frac{1}{{\tilde
      r}^3} - \frac{2 L_z^2}{{\tilde r}^4} \right],
\ee and \be {\tilde r}^4 \left( \frac{d \tilde \theta}{d \tilde t}
\right)^2 = K - \frac{L_z^2} {\sin^2 {\tilde\theta}} - Q E \cos(2
{\tilde \theta}). \label{s2q} \ee Note that the equations of motion
(\ref{s1q}) and (\ref{s2q}) have the same structure as the equations
of motion for Kerr geodesic motion. Using Eqs. (\ref{s1q}),
(\ref{s2q}) and (\ref{s3q}) together with the inverse of the
transformation (\ref{tildecoord}) to linear order in $Q$, we obtain
the expression for $K$ in spherical polar coordinates: \bea K &=&
r^4 ({\dot \theta}^2 + \sin^2\theta {\dot \varphi}^2)+ Q ({\dot r}
\cos\theta - r {\dot \theta}
\sin\theta )^2 + \frac{Q}{r}\nonumber\\
&-& \frac{Q}{2}({\dot r}^2 + r^2 {\dot \theta}^2 + r^2 \sin^2\theta
{\dot \varphi}^2) - \frac{2
  Q}{r} \cos^2\theta .\label{ksphericalq} \eea This is
equivalent to the formula (\ref{kcartesianq}) quoted earlier.

\subsection{Effects linear in spin on the conservative orbital
dynamics} To include the linear in spin effects, we repeat Ryan's
analysis \cite{ryan1,ryan2} (he only gives the final, time averaged
fluxes; we will also give the instantaneous fluxes). We can simply
add these linear in spin terms to our results because any terms of
order $O(SQ)$ will be higher than the order $a^2$ to which we are
working. The correction to the action (\ref{action}) due to
spin-orbit coupling is \be S^{\rm spin-orbit}=\int
dt\left[-\frac{2\mu Sn^i\epsilon_{ijk} x_j\dot
x_k}{r^3}\right].\label{soaction}\ee We will restrict our analysis
to the case when the unit vectors $n_i$ corresponding to the
axisymmetric quadrupole $Q_{ij}$ and to the spin $S_i$ coincide, as
they do in Kerr.

Including the spin-orbit term in the action (\ref{action}) results
in the following modified expressions for $L_z$ and $K$: \be L_z =
{\bf n} \cdot (\mu {\bf r} \times {\bf v}) -\frac{2S}{r^3}[{\bf
r}^2-({\bf n} \cdot {\bf r})^2],\label{lzcartesian} \ee and  \bea K
&=& ( {\bf r} \times {\bf v})^2 -\frac{4S}{r}{\bf n} \cdot ( {\bf r}
\times {\bf v})- \frac{2 Q }{r^3} ({\bf n} \cdot {\bf r})^2
\nonumber\\
&&+ Q \left[ ({\bf n} \cdot {\bf v})^2 - \frac{1}{2} {\bf v}^2 +
\frac{1}{r} \right]. \label{kcartesian}\eea In terms of the
Boyer-Lindquist like coordinates, the conserved quantities with the
linear in spin terms included are: \be L_z = {\tilde r}^2 \sin^2
{\tilde \theta} \left( \frac{d \varphi}{d{\tilde
      t}}\right) -\frac{2 S}{r}\sin^2 {\tilde \theta} -{Q} \sin^4 {\tilde
    \theta}\left( \frac{d \varphi}{d{\tilde
      t}}\right),
\label{s3} \ee \bea K &=& r^4 ({\dot \theta}^2 + \sin^2\theta {\dot
\varphi}^2) -4 S r \sin^2\theta \dot \varphi \nonumber\\&&- \frac{2
  Q}{r} \cos^2\theta
+  Q ({\dot r} \cos\theta - r {\dot \theta} \sin\theta )^2 +
\frac{Q M}{r}\nonumber\\
&& - \frac{Q}{2}({\dot r}^2 + r^2 {\dot \theta}^2 + r^2 \sin^2\theta
{\dot \varphi}^2).\label{kspherical} \eea The equations of motion
are \be \label{s1} \left( \frac{d \tilde r}{d \tilde t} \right)^2 =
2 E + \frac{2}{\tilde
  r} - \frac{K}{{\tilde r}^2} -\frac{4 S L_z}{{\tilde r}^3}+ \frac{Q}{2} \left[ \frac{1}{{\tilde
      r}^3} - \frac{2 L_z^2}{{\tilde r}^4} \right],
\ee and \be {\tilde r}^4 \left( \frac{d \tilde \theta}{d \tilde t}
\right)^2 = K - \frac{L_z^2} {\sin^2 {\tilde\theta}} - Q E \cos(2
{\tilde \theta}). \label{s2} \ee

\section{effects linear in quadrupole and quadratic in spin on the evolution of the constants of motion}
\label{sec2}

\subsection{Evaluation of the radiation reaction force}
The relative acceleration of the two bodies can be written as \be
{\bf a} = - {\bf \nabla} \Phi({\bf r}) + {\bf a}_{\rm rr}, \ee where
${\bf a}_{\rm rr}$ is the radiation-reaction acceleration. Combining
this with Eqs. (\ref{ecart}), (\ref{lzcartesian}) and
(\ref{kcartesian}) for $E$, $L_z$ and $K$ gives the following
formulae for the time derivatives of the conserved quantities:
\begin{eqnarray}
\label{cdot1}
{\dot E} &=&{\bf v} \cdot {\bf a}_{\rm rr}, \\
{\dot L_z} &=&  {\bf n} \cdot ({\bf r} \times {\bf a}_{\rm rr}), \\
{\dot K} &=& 2 ({\bf r} \times {\bf v}) \cdot ({\bf r} \times {\bf
a}_{\rm rr}) -\frac{ 4  S }{r}{\bf n} \cdot ({\bf r} \times {\bf
a}_{\rm rr})\nonumber\\
&& + 2 Q  ({\bf n} \cdot {\bf v}) \ ({\bf n} \cdot {\bf a}_{\rm rr}
) - Q {\bf v} \cdot {\bf a}_{\rm rr}. \label{cdot3}
\end{eqnarray}

The standard expression for the leading order radiation reaction
acceleration acting on one of the bodies is {\cite{bd}}: \bea
{a}^j_{\rm rr}&=& - \frac{2}{5}
I^{(5)}_{jk}x_k+\frac{16}{45}\epsilon_{jpq}S^{(6)}_{pk}x_kx_q+
\frac{32}{45}\epsilon_{jpq}S^{(5)}_{pk}x_kv_q\nonumber\\
&&+ \frac{32}{45}\epsilon_{pq[j}S^{(5)}_{k]p}x_qv_k. \label{arr}
\eea Here the superscripts in parentheses indicate the number of
time derivatives and square brackets on the indices denote
antisymmetrization.

The multipole moments $I_{jk}(t)$ and $S_{jk}(t)$ in Eq. (\ref{arr})
are the total multipole moments of the spacetime, i.e. approximately
those of the black hole plus those due to the orbital motion. The
expression (\ref{arr}) is formulated in asymptotically Cartesian
mass centered (ACMC) coordinates of the system, which are displaced
from the coordinates used in Sec. \ref{sec1} by an amount \cite{RMP}
\be \delta {\bf r}(t)=-\frac{\mu}{M}~{\bf r}(t).\label{acmcdispl}\ee
This displacement contributes to the radiation reaction acceleration
in the following ways:
\begin{enumerate}
\item The black hole multipole moments $I_l$ and $S_l$, which are
time-independent in the coordinates used in Sec. \ref{sec1}, will be
displaced by $\delta {\bf r}$ and thus will contribute to the
$(l+1)$th ACMC radiative multipole \cite{RMP}. \item The constants
of motion are defined in terms of the black hole centered
coordinates used in Sec. \ref{sec1}, so the acceleration ${\bf
a}_{\rm rr}$ we need in Eqs. (\ref{cdot1}) -- (\ref{cdot3}) is the
relative acceleration. This requires calculating the acceleration of
both the black hole and the point mass in the ACMC coordinates using
(\ref{arr}), and then subtracting to find ${\bf a}_{\rm rr}={\bf
a}^\mu_{\rm rr}-{\bf a}^M_{\rm rr}$ \cite{ryan1}. To leading order
in $\mu$, the only effect of the
 acceleration of the black hole is via a backreaction of the radiation field:
the $l$th black hole moments couple to the $(l+1)$th radiative
moments, thus producing an additional contribution to the
acceleration.\end{enumerate}

For our calculations at $O(S_1 \epsilon^3)$, $O(I_2\epsilon^4),$
$O(S_1^2\epsilon^4)$, we can make the following simplifications:

\begin{itemize}
\item {\it quadrupole corrections}: The fractional corrections linear
in $I_2=Q$ that scale as
 $O(a^2\epsilon^4)$ require only
the effect of $I_2$ on the conservative orbital dynamics as computed
in Sec. \ref{sec1}A and the Burke-Thorne
 formula for the radiation reaction acceleration [given by the first
 term in Eq. (\ref{arr})].
 \item {\it spin-spin corrections}: As discussed in the introduction, the
fractional corrections
 quadratic in $S_1$ to the conservative dynamics scale as $O(a^2\epsilon^6)$ and are subleading order effects which
 we neglect. At $O(a^2\epsilon^4)$, the only effect quadratic in $S_1$
 is the backscattering of the radiation off the spacetime
 curvature due to the spin. As discussed in item 1. above, the black hole's
 current dipole $S_i=S_1\delta_{i3}$ (taking the $z$-axis to be
 the symmetry axis) will contribute to the radiative
 current quadrupole an amount
\be S^{\rm spin}_{ij}=-\frac{3}{2}\frac{\mu}{M}S_1 x_i\delta_{j3}.
\label{sijspin} \ee The black hole's current dipole $S_i$ will
couple to the gravitomagnetic radiation field due to $S_{ij}$ as
discussed in item 2. above, and contribute to the relative
acceleration as \cite{ryan1}: \be {a}^{j{\;}{\rm spin}}_{\rm
rr}=\frac{8}{15}S_1\delta_{i3}S^{(5)}_{ij}. \label{arrspin}\ee For
our purposes of computing terms quadratic in the spin, we substitute
$S^{\rm spin}_{ij}$ for $S_{ij}$ in Eq. (\ref{arrspin}). Evaluating
these quadratic in spin terms requires only the Newtonian
conservative dynamics, i.e. the results of Sec. \ref{sec1} and Eqs.
(\ref{cdot1}) -- (\ref{cdot3}) with the quadrupole set to zero.
\item {\it linear in spin corrections}: Contributions to these effects are from Eq.
(\ref{arr}) with the current quadrupole replaced by just the spin
contribution (\ref{sijspin}), and from Eq. (\ref{arrspin}) evaluated
using only the orbital current quadrupole.
\end{itemize}

With these simplifications, we replace the expression (\ref{arr})
for the radiation reaction acceleration with \bea {a}^j_{\rm rr}&=&
- \frac{2}{5} I^{(5)}_{jk}x_k+\frac{16}{45}\epsilon_{jpq}S^{(6)~{\rm
spin}}_{pk}x_kx_q\nonumber\\
&&+ \frac{32}{45}\epsilon_{jpq}S^{(5)~{\rm spin}}_{pk}x_kv_q +
\frac{32}{45}\epsilon_{pq[j}S^{(5)~{\rm
spin}}_{k]p}x_qv_k\nonumber\\
&&+\frac{8}{15}S_1\delta_{i3}\left[S^{(5)~{\rm orbit}}_{ij}+
S^{(5)~{\rm spin}}_{ij}\right] . \label{btq} \eea

To justify these approximations, consider the scaling of the
contribution of black hole's acceleration to the orbital dynamics.
The mass and current multipoles of the black hole contribute terms
to the Hamiltonian that scale with $\epsilon$ as \be \Delta H \sim
S_l\epsilon^{2l+3}{\;}\&{\;}I_l\epsilon^{2l+2}. \ee Since the
Newtonian energy scales as $\epsilon^2$, the fractional correction
to the orbital dynamics scale as \be \Delta H/E\sim
S_l\epsilon^{2l+1}{\;}\&{\;}I_l\epsilon^{2l}.\label{orbitcorr} \ee
To $O(\epsilon^4),$ the only radiative multipole moments that
contribute to the acceleration (\ref{arr}) are the mass quadrupole
$I_2$, the mass octupole $I_3$, and the current quadrupole $S_2$
(cf. \cite{Kidder}). Since we are focusing only on the leading
order terms quadratic in spin (these can simply be added to the
known 2PN point particle and 1.5PN linear in spin results), the only
terms in Eq. $(\ref{arr})$ relevant for our purposes are those given
in Eq. $(\ref{btq}).$ The results from a computation of the fully
relativistic metric perturbation for black hole inspirals
\cite{pnkerr} show that quadratic in spin corrections to the $l=2$
piece compared to the flat space Burke-Thorne formula first appear
at $O(a^2\epsilon^4)$, which is consistent with the above arguments.

\subsection{Instantaneous fluxes}

 We evaluate the radiation reaction force as follows. The total mass and current quadrupole
 moment
 of the system are
 \bea
 \label{q}
Q^{\rm T}_{ij}&=&Q_{ij} + \mu x_i x_j, \\
S^{\rm T}_{ij}&=& S^{\rm spin}_{ij}+x_i \epsilon_{jkm}x_k\dot x_m,
 \label{s}
 \eea
  where from Eq. (\ref{tildecoord}) \bea x_i &=& \left[
{\tilde r} \sin{\tilde \theta} \left( 1 -
  \frac{Q}{4{\tilde r}^2} \right) \cos\varphi, ~
{\tilde r} \sin{\tilde \theta} \left( 1 -
  \frac{Q}{4{\tilde r}^2} \right) \sin\varphi,\right.\nonumber\\
&& ~ \left.{\tilde r} \cos{\tilde \theta} \left( 1 +
  \frac{Q}{4{\tilde r}^2} \right) \right].
\eea Only the second term in Eq. (\ref{q}) contributes to the time
derivative of the quadrupole. We differentiate five times by using
\be \frac{d}{d t} = \left[ 1 + \frac{Q}{2 {\tilde r}^2} \cos(2
{\tilde \theta}) \right] \frac{d}{d {\tilde t}}, \ee to the order we
are working as discussed above. After each differentiation, we
eliminate any occurrences of $d\varphi/d\tilde t$ using Eq.
(\ref{s3}), and we eliminate any occurrences of the second order
time derivatives $d^2\tilde r/d\tilde t^2$ and $d^2\tilde
\theta/d\tilde t^2$ in favor of first order time derivatives using
(the time derivatives of) Eqs. (\ref{s1}) and (\ref{s2}). For
computing the terms linear and quadratic in $S_1$, we set the
quadrupole $Q$ to zero in all the formulae. We insert the resulting
expression into the formula (\ref{btq}) for the self-acceleration,
and then into Eqs. (\ref{cdot1}) -- (\ref{cdot3}). We eliminate $(d{
\tilde r}/d\tilde t)^2$, $(d{ \tilde \theta}/d\tilde t)^2$, and $(d
\varphi/d\tilde t)$ in favor of $E$, $L_z$, and $K$ using Eqs.
(\ref{s3}) -- (\ref{s2}). In the final expressions for the
instantaneous fluxes, we keep only terms that are of $O(S)$, $O(Q)$
and $O(S^2)$ and obtain the following results:
\begin{widetext}
\noindent \bea
\dot E&=&\frac{160 K}{3 {\tilde r}^6} +\frac{64}{3{\tilde r}^5}+\frac{512 E}{15
{\tilde r}^4}-\frac{40
K^2}{{\tilde r}^7}+\frac{272KE}{5{\tilde r}^5}+\frac{64E^2}{5{\tilde r}^3}\nonumber\\
&+&\frac{SL_z}{{\tilde r}^9}\left(196 K^2+\frac{952}{3}{\tilde r}^2-\frac{3668}{5}
K{\tilde r}-352 KE{\tilde r}^2+\frac{1024}{3}E{\tilde r}^3+\frac{128}{5}E^2
{\tilde r}^4\right)\nonumber\\
&+&\frac{2Q}{{\tilde r}^9}\left[-49 K^2
-169KL_z^2+{\tilde r}\left(\frac{532}{5}K+\frac{3307}{15}L_z^2\right)
+2{\tilde r}^2\left(-\frac{20}{3}+47KE+\frac{548}{5}L_z^2E\right)-\frac{152}{5}{\tilde r}^3E-16
{\tilde r}^4E^2\right]\nonumber\\
&+&\frac{Q}{{\tilde r}^9}\left[\left(-562K^2+\frac{2998}{3}K{\tilde r}-\frac{320}{3}{\tilde r}^2+\frac{5072}{5}KE{\tilde r}^2-\frac{4048}{15}{\tilde r}^3E-
160 {\tilde r}^4E^2\right)\cos (2 {\tilde \theta})\right]\nonumber\\
&+& \frac{Q}{{\tilde r}^6}\sin (2
{\tilde \theta})\left(439K-\frac{926}{3}{\tilde r}-\frac{1528}{5}{\tilde r}^2E\right)\dot{\tilde \theta}\dot
{\tilde r} \nonumber\\
&+&\frac{S^2}{{\tilde r}^9}\left[\left(-K^2+\frac{22}{3}K{\tilde r}-\frac{28}{3}{\tilde r}^2+\frac{32}{5}KE{\tilde r}^2-\frac{236}{15}{\tilde r}^3E
-\frac{32}{5} {\tilde r}^4E^2\right)\cos (2 {\tilde \theta}) -{\tilde r}^3\sin (2
{\tilde \theta})\left(K+\frac{2}{3}{\tilde r}+\frac{8}{5}{\tilde r}^2E\right)\dot{\tilde \theta}\dot
{\tilde r}\right] \nonumber\\
&+& \frac{S^2}{{\tilde r}^9}\left[-49 K^2 +6
KL_z^2+2{\tilde r}\left(63K-\frac{16}{3}L_z^2-\frac{98}{3}\right)
+{\tilde r}^2\left(112KE-\frac{48}{5}L_z^2E\right)-\frac{1652}{15}{\tilde r}^3E-\frac{224}{5}
{\tilde r}^4E^2\right], \label{edotinst}\eea
\noindent \bea \dot L_z&=&\frac{32L_z}{{\tilde r}^4}+\frac{144L_zE}{5{\tilde r}^3}-\frac{24KL_z}{{\tilde r}^5}\nonumber\\
&+&\frac{S}{{\tilde r}^7}\left[-50K^2+240KL^2_z+\frac{62}{5}K{\tilde r}-\frac{7376}{15}L^2_z
{\tilde r}+\frac{316}{3}{\tilde r}^2+56KE{\tilde r}^2-\frac{1824}{5}EL^2_z{\tilde r}^2+\frac{624}{5}E{\tilde r}^3+\frac{128}{5}E^2{\tilde r}^4\right]\nonumber\\
&+&\frac{S}{{\tilde r}^7}\left(50K^2-\frac{62}{5}K{\tilde r}-\frac{316}{3}{\tilde r}^2-56KE{\tilde r}^2-\frac{624}{5}E{\tilde r}^3-\frac{128}{5}E^2{\tilde r}^4\right)
\cos (2{\tilde \theta})\nonumber\\
&+&\frac{S}{{\tilde r}^4}\left(-104K+64{\tilde r}+64 E{\tilde r}^2\right)\sin (2{\tilde \theta})\dot
{\tilde r}\dot {\tilde \theta}\nonumber\\
&+&\frac{QL_z}{5{\tilde r}^7}\left[660E{\tilde r}^2+753{\tilde r}-360L_z^2-435K+\left(1601{\tilde r}+1512{\tilde r}^2E-1185K\right)\cos
2{\tilde \theta}\right]+\frac{174QL_z}{{\tilde r}^4}\sin (2{\tilde \theta})\dot {\tilde r}\dot
{\tilde \theta}\nonumber\\
&+&\frac{2S^2L_z}{{\tilde r}^7}\left[\frac{72}{5}E{\tilde r}^2+16{\tilde r}-9K\right],\label{lzdotinst}\eea
and
\noindent \bea \dot K&=&\frac{16K}{5{\tilde r}^5}\left(20{\tilde r}+18{\tilde r}^2
E-15K\right)\nonumber\\
&+&\frac{SL_z}{{\tilde r}^7}\left(280K^2-\frac{14008}{15}K
{\tilde r}+\frac{1264}{3}{\tilde r}^2+\frac{2496}{5}E
{\tilde r}^3-\frac{2528}{5}KE{\tilde r}^2+\frac{512}{5}E^2{\tilde r}^4\right)\nonumber\\
&+&\frac{12 Q}{5 \tilde r^7}\left[-45 K^2+\tilde r L_z^2(83+80\tilde r
  E)-115 K L_z^2+14K\tilde r (6+5 \tilde r E)\right]\nonumber\\
&+&\frac{4 Q}{15 {\tilde r}^7} \cos
(2{\tilde \theta})\left(-2175K^2+2975K{\tilde r}+80{\tilde r}^2+3012KE{\tilde r}^2-112E{\tilde r}^3-168E^2{\tilde r}^4\right)\nonumber\\
&+&\frac{2 Q}{15 {\tilde r}^4}\left(3075K-20{\tilde r}-192E{\tilde r}^2\right)\sin
(2{\tilde \theta})\dot{\tilde \theta}\dot
{\tilde r} \nonumber\\
&+&\frac{2S^2}{
{\tilde r}^7}\left[\left(7K-2L_z^2\right)\left(-3K+\frac{16}{3}{\tilde r}+\frac{24}{5}E{\tilde r}^2\right)
+K\cos
(2{\tilde \theta})\left(3K-\frac{16}{3}{\tilde r}-\frac{24}{5}E{\tilde r}^2\right)\right]\nonumber\\
&+&\frac{2S^2}{ {\tilde r}^4}\sin
(2{\tilde \theta})\left(-4K+\frac{14}{3}{\tilde r}+\frac{16}{5}E{\tilde r}^2\right)\dot{\tilde \theta}\dot
{\tilde r} .\label{kdotinst}\eea
\end{widetext}

\subsection{Alternative set of constants of the motion}

A body in a generic bound orbit in Kerr traces an open ellipse
precessing about the hole's spin axis. For stable orbits the motion
is confined to a toroidal region whose shape is determined by $E$,
$L_z$, $K$. The motion can equivalently be characterized by the set
of constants inclination angle $\iota$, eccentricity $e$, and
semi-latus rectum $p$ defined by Hughes \cite{hughes}. The constants
$\iota$, $p$ and $e$ are defined by $ \cos\iota=L_z/ \sqrt{K}$, and
by $\tilde r_\pm = p / (1 \pm e)$, where $\tilde r_\pm$ are the
turning points of the radial motion, and $\tilde r$ is the
Boyer-Lindquist radial coordinate. This parameterization has a simple
physical interpretation: in the Newtonian limit of large $p$, the
orbit of the particle is an ellipse of eccentricity $e$ and
semilatus rectum $p$ on a plane whose inclination angle to the
hole's equatorial plane is $\iota$. In the relativistic regime
$p\sim M$, this interpretation of the constants $e$, $p$, and
$\iota$ is no longer valid because the orbit is not an ellipse and
$\iota$ is not the angle at which the object crosses the equatorial
plane (see Ryan \cite{ryan1} for a discussion).

We adopt here analogous definitions of constants of motion $\iota$,
$e$ and $p$, namely \bea \label{iotadef} \cos(\iota)&=&L_z/\sqrt{K},\\
\frac{p}{1\pm e}&=&\tilde r_\pm . \label{epdef}\eea Here $K$ is the
conserved quantity (\ref{kcartesian}) or (\ref{kspherical}), and
$\tilde r_\pm$ are the turning points of the radial motion using the
$\tilde r$ coordinate defined by Eq. (\ref{tildecoord}), given by
the vanishing of the right-hand side of Eq. (\ref{s1}).

We now rewrite our results in terms of the new constants of the
motion $e$, $p$ and $\iota$. We can use Eq. (\ref{s1}) together with
the equations (\ref{iotadef}) and (\ref{epdef}) to write $E$, $L_z$
and $K$ as functions of $ p$, $ e$ and $\iota$. To leading order in
$Q$ and $S$ we obtain \bea \label{krelation} K&=&
p\left[1-\frac{2S\cos\iota}{p^{3/2}}\left(3+e^2\right)
-\left(1+ e^2\right)\frac{2Q\cos^2\iota}{p^2}\right.\nonumber\\
&& ~ ~ \left.+\left(3+e^2\right)\frac{Q}{4p^2}\right],\\
E&=&-\frac{(1-e^2)}{2p}\left[1+\frac{2S\cos\iota}{p^{3/2}}\left(1-e^2\right)\right.\nonumber\\
&& {\;}{\;}{\;}{\;}{\;}{\;}{\;}{\;}{\;}
{\;}{\;}{\;}{\;}{\;}{\;}{\;}{\;}
\left.+\left(1-e^2\right)\frac{Q}{p^2}\left(\cos^2\iota-\frac{1}{4}\right)\right],\\
L_z&=&\sqrt{p}\cos \iota \left[1-\frac{S\cos\iota}{p^{3/2}}(3+e^2)
-\left(1+e^2\right)\frac{Q\cos^2\iota}{p^2}\right.\nonumber\\
&& ~ ~
{\;}{\;}{\;}{\;}{\;}{\;}{\;}{\;}{\;}{\;}{\;}{\;}\left.+\left(3+e^2\right)\frac{Q}{8p^2}\right].\label{orbitalelem}\eea
As discussed in the introduction, the effects quadratic in $S$ on
the conservative dynamics scale as $O(a^2\epsilon^6)$ and thus are
not included in this analysis to $O(a^2\epsilon^4)$.

Inserting these relations into the expressions
(\ref{edotinst})--(\ref{kdotinst}) gives, dropping terms of $O(QS)$,
$O(Q^2)$ and $O(QS^2)$: \begin{widetext} \bea \dot
E&=&-\frac{8}{15p^2 {\tilde r}^7}\left[75 p^4-100p^3{\tilde r}+p^2{\tilde r}^2\left(11-51
e^2\right)+32p{\tilde r}^3\left(1-e^2\right))-6{\tilde r}^4\left(1-e^2\right)^2\right]\nonumber\\
&+&\frac{4S\cos\iota}{15p^{7/2}{\tilde r}^9}\left[735p^6-2751p^5{\tilde r}+10p^4{\tilde r}^2(365-6e^2)-128p{\tilde r}^5(1-e^2)^2-48{\tilde r}^6(e^2-1)^3\right]\nonumber\\
&+&\frac{64 S\cos\iota}{15p^{3/2}{\tilde r}^6}\left[5p(-23+3e^2)-3{\tilde r}(-9+e^2+8e^4)\right]\nonumber\\
&-&\frac{Q}{15p^4{\tilde r}^9}\left[4005p^6-6499p^5{\tilde r}+2p^4{\tilde r}^2\left(1577-1977
e^2\right)-24{\tilde r}^6\left(1-e^2\right)^3-32p^3{\tilde r}^3\left(8-33e^2\right)+64p{\tilde r}^5\left(1-2e^2+e^4\right)\right]\nonumber\\
&-&\frac{Q}{15p^4{\tilde r}^9}\left[24 p^2{\tilde r}^4\left(5-27 e^2+22e^4\right)
-p{\tilde r}^3\sin (2{\tilde \theta})\left(6585p^2-4630p{\tilde r}+2292{\tilde r}^2(1-e^2)\right)
\dot{\tilde \theta} \dot {\tilde r}\right]\nonumber\\
&-&\frac{Q}{15p^4{\tilde r}^9}\left[2p^2\cos (2{\tilde \theta})\left(4215p^4-7495p^3
{\tilde r}+4p^2{\tilde r}^2(1151-951e^2)-1012p{\tilde r}^3(1-e^2)+300{\tilde r}^4(1-2e^2+e^4)\right)\right]\nonumber\\
&-&\frac{Q}{15p^4{\tilde r}^9}\cos
(2\iota)\left[2535p^6-3307p^5{\tilde r}+12p^4{\tilde r}^2(37-237e^2)-48{\tilde r}^6(1-e^2)^3
+800p^3{\tilde r}^3(1+e^2)+128p{\tilde r}^5(1-2e^2+e^4)\right]\nonumber\\
&+&\frac{204Q}{15p^2{\tilde r}^5}\cos
(2\iota)\left(1+2e^2-3e^4\right)\nonumber\\
&-&\frac{2S^2}{15p^2 {\tilde r}^9
}\left[84{\tilde r}^4(1-e^2)^2(1+e^2)^2+345p^4-905p^3{\tilde r}-413p{\tilde r}^3(1-e^2)+2p^2{\tilde r}^2(446-201e^2)\right]\nonumber\\
&-&\frac{S^2}{15 p^2{\tilde r}^9}\cos (2{\tilde \theta}) \left[15
p^4-110p^3{\tilde r}+4p^2{\tilde r}^2(47-12e^2)-118p{\tilde r}^3(1-e^2)+24{\tilde r}^4(1-e^2)^2(1+e^2)^2\right]\nonumber\\
&+&\frac{S^2}{15 {\tilde r}^9}\cos
(2\iota)\left[45p^2-80p{\tilde r}+36{\tilde r}^2(1-e^2)\right]-\frac{S^2}{15p{\tilde r}^6}\sin
(2{\tilde \theta})\dot {\tilde r}\dot{\tilde \theta}\left[15p^2+10p{\tilde r}-12{\tilde r}^2(1-e^2)\right],
\label{Edotinst} \eea
\noindent \bea \dot
L_z&=&-\frac{8\cos\iota}{5\sqrt{p}{\tilde r}^5}\left[15p^2-20
p{\tilde r}+9{\tilde r}^2(1-e^2)\right]\nonumber\\
&+&\frac{2S}{15p^2{\tilde r}^7}\left[525p^4-1751p^3{\tilde r}+34p^2{\tilde r}^2(61-6e^2)+12p{\tilde r}^3(-69+29e^2)+6{\tilde r}^4(17+2e^2-19e^4)\right]\nonumber\\
&+&\frac{2S}{15p^2{\tilde r}^7}\left[375p^4-93p^3{\tilde r}+468p{\tilde r}^3(1-e^2)-10p^2{\tilde r}^2(58+21e^2)-48{\tilde r}^4(1-2e^2+e^4)\right]\cos(2{\tilde \theta})\nonumber\\
&+&\frac{4S}{15p^2{\tilde r}^7}\left[450p^4-922p^3{\tilde r}-60p{\tilde r}^3(3+e^2)-9p^2{\tilde r}^2(-83+23e^2)+27{\tilde r}^4(1+2e^2-3e^4)\right)\cos(2\iota)\nonumber\\
&-&\frac{8S}{p{\tilde r}^4}\left[13p^2-8p{\tilde r}+4{\tilde r}^2(1-e^2)\right]\sin(2{\tilde \theta})\dot
{\tilde r} \dot {\tilde \theta}\nonumber\\
&-&\frac{Q\cos\iota}{5
p^{5/2}{\tilde r}^7}\left[615p^4-753p^3{\tilde r}+15p^2{\tilde r}^2\left(19-31e^2\right)+20
p{\tilde r}^3\left(1+3e^2\right)+9{\tilde r}^4\left(1-6e^2+5e^4\right)\right]\nonumber\\
&-&\frac{Q\cos\iota}{5p^{1/2}{\tilde r}^7}\cos (2{\tilde \theta})\left(1185p^2-1601p
{\tilde r}+756{\tilde r}^2(1-e^2)\right)\nonumber\\
&-&\frac{2Q\cos\iota}{5p^{5/2}{\tilde r}^7}\left[2 \cos
(2\iota)\left(45p^4-18{\tilde r}^4e^2(1-e^2)-45p^2{\tilde r}^2(1+e^2)+20p{\tilde r}^3(1+e^2)\right)-435p^3{\tilde r}^3\sin
(2 {\tilde \theta}) \dot {\tilde \theta} \dot {\tilde r}\right]\nonumber\\
&-&\frac{2S^2\cos\iota}{p^{1/2}{\tilde r}^7}\left[9p^2-16p{\tilde r}+\frac{36}{5}{\tilde r}^2(1-e^2)\right],
\label{Lzdotinst}\eea and
\noindent\bea \dot K&=&\frac{16}{5{\tilde r}^5}\left[20p{\tilde r}-15p^2-9{\tilde r}^2(1-e^2)\right]\nonumber\\
&+&\frac{8S\cos
\iota}{15p^{3/2}{\tilde r}^7}\left[525p^4-1751p^3{\tilde r}+2p^2{\tilde r}^2(1172-57e^2)+12p{\tilde r}^3(-99+19e^2)-24{\tilde r}^4(-11+4e^2+7e^4)\right]\nonumber\\
&+&\frac{2Q}{5p^2\tilde r^7} \left[-615p^4+753p^3\tilde r+30 p^2\tilde
  r^2 (17 e^2-9)+72\tilde r^4e^2(1-e^2)-40p\tilde r^3 (1+3e^2)\right]\nonumber\\
&+& \frac{2Q}{5p^2\tilde r^7}\cos(2\iota)\left[-345 p^4+249 p^3\tilde
  r-160p\tilde r^3(1+e^2)+120 p^2\tilde r^2(1+3e^2)+36\tilde r^4(1+2e^2-3e^4)\right]\nonumber\\
&+& \frac{2Q}{15 p^2 {\tilde r}^7}\left[2\cos
(2{\tilde \theta})\left(2175p^4-2975p^3{\tilde r}-56p{\tilde r}^3(1-e^2)+2p^2{\tilde r}^2(713-753e^2)+42{\tilde r}^4(1-2e^2+e^4)\right)\right]\nonumber\\
&+& \frac{2Q}{15 p {\tilde r}^4}\sin (2{\tilde \theta})
\left(3075p^2-20p{\tilde r}+96{\tilde r}^2(1-e^2)\right)\dot {\tilde r}\dot {\tilde \theta}\nonumber\\
&+&
\frac{2S^2}{{\tilde r}^7}\left\{2\left[-9p^2+16p{\tilde r}-\frac{36}{5}{\tilde r}^2(1-e^2)\right]+\left(\cos
(2{\tilde \theta})+\cos (2\iota)\right)\left[3p^2-\frac{16}{3}p{\tilde r}+\frac{12}{5}
{\tilde r}^2(1-e^2)\right]\right\}\nonumber\\
&+& \frac{4S^2}{p{\tilde r}^4}\sin (2{\tilde \theta})\dot {\tilde r}\dot
{\tilde \theta}\left[-2p^2+\frac{7}{3}p{\tilde r}-\frac{4}{5}{\tilde r}^2(1-e^2)\right].
\label{Kdotinst}\eea
\end{widetext}

\subsection{Time averaged fluxes} In this section we will compute
the infinite time-averages $\langle \dot E\rangle$, $\langle \dot
L_z\rangle$ and $\langle\dot K\rangle$ of the fluxes. These averages
are defined by \be \langle\dot E\rangle\equiv
\lim_{T\to\infty}\frac{1}{T}\int^{T/2}_{-T/2}\dot E(t) dt. \ee These
time-averaged fluxes are sufficient to evolve orbits in the
adiabatic regime (except for the effect of resonances)
\cite{Mino2003,res}. In Appendix II, we present two different ways of
computing the time averages. The first approach is based on
decoupling the $\tilde r$ and $\tilde \theta$ motion using the
analog of the Mino time parameter for geodesic motion in Kerr
\cite{Mino2003}. The second approach uses the explicit Newtonian
parameterization of the orbital motion. Both averaging methods give
the following results:
\begin{widetext}
\noindent
\bea \langle \dot E\rangle
&=&-\frac{32}{5}\frac{(1-e^2)^{3/2}}{p^5}\left[1+\frac{73}{24}e^2+\frac{37}{96}e^4
-\frac{S}{p^{3/2}}\left(\frac{73}{12}+\frac{823}{24}e^2+\frac{949}{32}e^4+\frac{491}{192}e^6\right)
\cos(\iota)\right.\nonumber\\
&&{\;}{\;}{\;}{\;}{\;}{\;}{\;}{\;}{\;}{\;}{\;}{\;}{\;}\left.-\frac{Q}{p^2}\left\{\frac{1}{2}+\frac{85}{32}e^2+\frac{349}{128}e^4+\frac{107}{384}e^6
+\left(\frac{11}{4}+\frac{273}{16}e^2+\frac{847}{64}e^4+\frac{179}{192}e^6\right)\cos
(2 \iota)\right\}\right.\nonumber\\
&&{\;}{\;}{\;}{\;}{\;}{\;}{\;}{\;}{\;}{\;}{\;}{\;}{\;}+\left.\frac{S^2}{p^2}\left\{\frac{13}{192}+\frac{247}{384}e^2+\frac{299}{512}e^4+\frac{39}{1024}e^6
-\left(\frac{1}{192}+\frac{19}{384}e^2+\frac{23}{512}e^4+\frac{3}{1024}e^6\right)\cos
(2 \iota)\right\}\right],\label{edotavg}\eea

\noindent \bea \langle \dot L_z\rangle
&=&-\frac{32}{5}\frac{(1-e^2)^{3/2}}{p^{7/2}}\cos
\iota\left[1+\frac{7}{8}e^2-\frac{S}{2
p^{3/2}\cos\iota}\left\{\frac{61}{24}+7e^2+\frac{271}{64}e^4+\left(\frac{61}{8}+\frac{91}{4}e^2+\frac{461}{64}e^4\right)
\cos(2\iota)\right\}\right.\nonumber\\
&&{\;}{\;}{\;}{\;}{\;}{\;}{\;}{\;}{\;}{\;}{\;}{\;}{\;}{\;}{\;}{\;}{\;}{\;}{\;}{\;}{\;}{\;}{\;}{\;}{\;}
{\;}{\;}{\;}{\;}{\;}{\;}{\;}{\;}{\;}{\;}
\left.-\frac{Q}{16p^2}\left\{-3-\frac{45}{4}e^2+\frac{19}{8}e^4
+\left(45+148e^2+\frac{331}{8}e^4\right)\cos
(2\iota)\right\}\right.\nonumber\\
&&{\;}{\;}{\;}{\;}{\;}{\;}{\;}{\;}{\;}{\;}{\;}{\;}{\;}{\;}{\;}{\;}{\;}{\;}{\;}{\;}{\;}{\;}{\;}{\;}{\;}
{\;}{\;}{\;}{\;}{\;}{\;}{\;}{\;}{\;}{\;}
\left.+\frac{S^2}{16p^2}\left\{1+3e^2+\frac{3}{8}e^4\right\}\right],\label{lzdotavg}\eea
\noindent \bea \langle \dot K\rangle
&=&-\frac{64}{5}\frac{(1-e^2)^{3/2}}{p^{3}}\left[1+\frac{7}{8}e^2
-\frac{S}{2p^{3/2}}\left(\frac{97}{6}+37e^2+\frac{211}{16}e^4\right)\cos(\iota)\right.\nonumber\\
&&{\;}{\;}{\;}{\;}{\;}{\;}{\;}{\;}{\;}{\;}{\;}{\;}{\;}{\;}{\;}{\;}{\;}{\;}{\;}{\;}{\;}{\;}{\;}{\;}{\;}{\;}{\;}{\;}
\left.-\frac{Q}{p^2}\left\{1+
\frac{8}{3}e^2+\frac{11}{12}e^4+\left(\frac{13}{4}+\frac{841}{96}e^2+\frac{449}{192}e^4\right)\cos
(2\iota)\right\}\right.\nonumber\\
&&{\;}{\;}{\;}{\;}{\;}{\;}{\;}{\;}{\;}{\;}{\;}{\;}{\;}{\;}{\;}{\;}{\;}{\;}{\;}{\;}{\;}{\;}{\;}{\;}{\;}{\;}{\;}{\;}
\left.+\frac{S^2}{p^2}\left\{\frac{13}{192}+
\frac{13}{64}e^2+\frac{13}{512}e^4-\left(\frac{1}{192}+\frac{1}{64}e^2+\frac{1}{512}e^4\right)\cos
(2\iota)\right\}\right].\label{kdotavg} \eea

 Using Eqs. (\ref{krelation}) and (\ref{orbitalelem}), we obtain
 from (\ref{edotavg}) -- (\ref{kdotavg}) the following time averaged rates of change of the orbital
elements $e,$ $p,$ $\iota$: 
\begin{eqnarray}
 \langle \dot p\rangle &=& -\frac{64}{5}\frac{(1-e^2)^{3/2}}{p^3}\left\{1+\frac{7e^2}{8}-\frac{S\cos(\iota)}{96p^{3/2}}\left(1064+1516e^2+475e^4\right)\right.\nonumber\\
&& \ \ \ \ \ \ \ \ \ \ \ \ -\frac{Q}{8p^{2}}\left[14+\frac{149e^2}{12}+\frac{19e^4}{48}+\left(50+\frac{469e^2}{12}+\frac{227 e^4}{24}\right)\cos(2\iota)\right]\nonumber\\
&&\ \ \ \ \ \ \ \   \ \ \ +\left.\frac{S^2}{64 p^2}\left(\frac{1}{3}+e^2+\frac{e^4}{8}\right)\left[13-\cos(2\iota)\right]\right\}, \ \ \ \ \ \ \ \ \ \  \ \ \ \ \ \ \ \ \ \ \ \ \ \label{pdot}
\end{eqnarray}

\begin{eqnarray}
 \langle\dot e\rangle&=& -\frac{304}{15}\frac{e(1-e^2)^{3/2}}{p^4}\left(1+\frac{121e^2}{304}\right)+\frac{Se(1-e^2)^{3/2}\cos(\iota)}{5 p^{11/2}}\left(1172+932e^2+\frac{1313e^4}{6}\right)\nonumber\\
&& +\frac{Q(1-e^2)^{3/2}}{ ep^6}\left[32+\frac{785 e^2}{3}-\frac{219e^4}{2}+13e^6+\left(32+\frac{2195e^2}{3}+251e^4+\frac{218e^6}{3}\right)\cos(2\iota)\right]\nonumber\\
&&-\frac{S^2e(1-e^2)^{3/2}}{8 p^6}\left(2+3e^2+\frac{e^4}{4}\right)\left[13-\cos(2 \iota)\right],\label{edot}
\end{eqnarray}

\begin{eqnarray}
  \langle \dot \iota\rangle &=& \frac{S\sin (\iota) (1-e^2)^{3/2}}{ p^{11/2}}\left[\frac{244}{15}+\frac{252}{5}e^2+\frac{19}{2}e^4\right]-\frac{\left(1-e^2\right)^{3/2}  S^2  \sin (2 \iota )}{240
   p^6}\left[8+3 e^2\left(8+e^2\right)\right]\nonumber\\
&&+\frac{Q\cot(\iota)(1-e^2)^{3/2}}{60p^{6}}\left[312+736 e^2-83 e^4-\left(408+1268 e^2+599e^4\right)\cos(2\iota)\right].\label{iotadot}
\end{eqnarray}

\end{widetext}

\section{application to black holes}
\label{sec3}

\subsection{Qualitative discussion of results}
The above results for the fluxes, Eqs. (\ref{pdot}), (\ref{edot})
and (\ref{iotadot}) show that the correction terms at
$O(a^2\epsilon^4)$ due to the quadrupole have the same type of
effect on the evolution as the linear spin correction computed by
Ryan: they tend to circularize eccentric orbits and change the angle
$\iota$ such as to become antialigned with the symmetry axis of the
quadrupole.

The effects of the terms quadratic in spin are qualitatively
different. In the expression (\ref{edotavg}) for $\langle \dot
E\rangle$, the coefficient of $\cos(2\iota)$ due to the spin
self-interaction has the same sign as the quadrupole term, while
the terms not involving $\iota$ have the opposite sign. The terms
involving $\cos(2\iota)$ in Eq. (\ref{kdotavg}) for $\langle \dot
K\rangle$ of $O(Q)$ and $O(S^2)$ terms have the same sign, while the
terms not involving $\iota$ have the opposite sign. The fractional
spin-spin correction to $\langle \dot L_z\rangle $, Eq.
(\ref{lzdotavg}), has no $\iota$-dependence, and in expression
(\ref{iotadot}) for $\langle \dot \iota\rangle$, the dependence on
$\iota$ of the two effects $O(Q)$ and $O(S^2)$ is different, too.
This is not surprising as the $O(Q)$ effects included here are
corrections to the conservative orbital dynamics, while the effects
of $O(S^2)$ that we included are due to radiation reaction.

\subsection{Comparison with previous results}
The terms linear in the spin in our results for the time averaged
fluxes, Eqs. (\ref{edotavg}) -- (\ref{iotadot}), agree with those
computed by Ryan, Eqs. (14a) -- (15c) of \cite{ryan2}, and with
those given in Eqs. (2.5) -- (2.7) of Ref. \cite{ghk}, when we use
the transformations to the variables used by Ryan given in Eqs.
(2.3) -- (2.4) in \cite{ghk}.

Equation (\ref{edotavg}) for the time averaged energy flux agrees
with Eq. (3.10) of Gergely \cite{Gergely} and Eq. (4.15) of
\cite{gergelyspin} when we use the following transformations: \bea
\label{gergktranf} K&=& {\bar L}^2\left[1-\frac{Q}{2{\bar
L}^4}\left({\bar A}^2\sin^2\kappa\cos\delta-(1-{\bar
A}^2)\cos^2\kappa\right)\right]\nonumber\\
&=&{\bar L}^2\left[1-\frac{Q}{{\bar
L}^4} ~ E ~ \cos^2\kappa\right.\nonumber\\
&&{\;}{\;}{\;}{\;}{\;}{\;}{\;}{\;}\left.-\frac{Q}{2{\bar
L}^4}(1+2{\bar L}^2)\sin^2\kappa\cos\delta\right],\\
\cos\iota &=& \cos\kappa \left[1+\frac{Q}{2{\bar L}^4} ~ E ~
\cos^2\kappa\right.\nonumber\\
&&{\;}{\;}{\;}{\;}{\;}{\;}{\;}{\;}{\;}{\;}{\;}{\;}\left.+\frac{Q}{2{\bar
L}^4}(1+2{\bar
L}^2)\sin^2\kappa\cos\delta\right],\\
\xi_0&=&\frac{1}{2}(\delta+\kappa),\\
\xi_0&=&\left(\psi_0-\psi_i\right)+\frac{\pi}{2},
\label{gergtransf}\eea where $\bar A$, $\bar L$, $\kappa$, $\delta$,
$\psi_0$ and $\psi_i$ are the quantities used by Gergely. The first
relation here is obtained from the turning points of the radial
motion as follows. We compute $\tilde r_\pm$ in terms of $E$ and $K$
and map these expressions back to $r$ using Eqs. (\ref{tildecoord}).
The result can then be compared with the turning points in Gergely's
variables, Eq. (2.19) of \cite{Gergely}, using the fact that $E$ is
the same in both cases. Instead of the evolution of the constants of
motion $K$ and $L_z$, Gergely computes the rates of change of the
magnitude $L$ of the orbital angular momentum and of the angle
$\kappa$ defined by $\cos\kappa=({\bf L}\cdot {\bf S})/L$. Using the
transformations (\ref{gergktranf}) -- (\ref{gergtransf}) and the
definition of $\kappa$ we verify that our Eq. (\ref{lzdotavg})
agrees with the $\langle \dot L_z\rangle$ computed using Gergely's
Eqs. (3.23) and (3.35) in \cite{Gergely} and Eq. (4.30) of
\cite{gergelyspin}.

In the limit of the circular equatorial orbits analyzed by Poisson
\cite{poisson}, our Eq. (\ref{edotavg}) agrees with Poisson's Eq.
(22) when we use the transformations and specializations: \bea
\label{circtransf}
p&=&\frac{1}{v^2}\left[1-\frac{Q}{4}v^4\right],\\
\iota&=&0,\\
e^2&=&0,\\
\cos\alpha_A&=&1, \eea where $v$ and $\alpha_A$ are the variables
used by Poisson and the relation (\ref{circtransf}) is obtained by
comparing the expressions for the constants of motion in the two
sets of variables.

The main improvement of our analysis over Gergely's is that we
express the results in terms of the Carter-type constant $K$, which
facilitates comparing our results with other analyses of black hole
inspirals. Our computations also include the spin curvature
scattering effects for all three constants of motion; Gergely
\cite{gergelyspin} only considers these effects for two of them: the
energy and magnitude of angular momentum, not for the third
conserved quantity.

When we expand Eq. (\ref{edotavg}) for small inclination angles and
specialize to circular orbits, then after converting $p$ to the
parameter $v$ using Eq. (\ref{circtransf}), we obtain \bea \langle
\dot E\rangle &=&-\frac{32}{5p^5}\left[1-\frac{1}{p^2}\left(2
Q+\frac{S^2}{16}\right)+\frac{\iota^2}{2p^2}\left(11
Q-\frac{S^2}{48}\right)\right]\nonumber\\
&=& -\frac{32}{5p^5}\left[1-\frac{a^2
v^4}{16}\left\{33-\frac{527}{6}\iota^2\right\}\right].\label{edotshib}\eea
This result agrees with the terms at $O(a^2 v^4)$ of Eq. (3.13) of
Shibata et al. \cite{pnkerr}, whose calculations were based on the
fully relativistic expressions. This agreement is a check that we
have taken into account all the contributions at $O(a^2\epsilon^4)$.
The analysis in Ref. \cite{pnkerr} could not distinguish between
effects due to the quadrupole and those due curvature scattering,
but we can see from Eq. (\ref{edotshib}) that those two interactions
have the opposite dependence on $\iota$. Comparing (\ref{edotshib})
with Eq. (3.7) of \cite{pnkerr} (which gives the fluxes into the
different modes $(l=2,m,n)$, where $m$ and $n$ are the multiples of
the $\varphi$ and $\theta$ frequencies), we see that the terms in
the $(2,\pm 2,0)$ and the $(2,\pm 1, \pm 1)$ modes are entirely due
to the quadrupole, while the spin-spin interaction effects are fully
contained in the $(2, \pm 1, 0)$ and $(2,0,\pm 1)$ modes.

\section{non-existence of a Carter-type constant for higher multipoles}
\label{sec4}

In this section, we
show that for a single axisymmetric multipole interaction, it is not
possible to find an analog of the Carter constant (a conserved
quantity which does not correspond to a symmetry of the Lagrangian),
except for the cases
of spin (treated by Ryan \cite{ryan2})
and mass quadrupole moment
(treated in this paper). Our proof is valid only in the approximations
in which we work -- expanding to linear order in the mass ratio, to
the leading
post-Newtonian order, and to linear order in the multipole.  However
we will show below that with very mild additional smoothness
assumptions, our non-existence result extends to exact geodesic motion
in exact vacuum spacetimes.

We start in Sec.\ \ref{separability} by showing that
there is no coordinate system in which the Hamilton-Jacobi equation is
separable.  Now separability of the Hamilton-Jacobi equation is a
sufficient but not a necessary condition for the existence of a
additional conserved quantity.  Hence, this result does not yield
information about the existence or non-existence of an additional
constant.  Nevertheless we find it to be a suggestive result.
Our actual derivation of the non-existence is based on Poisson bracket
computations, and is given in Sec.\ \ref{poisson}.

\subsection{Separability analysis}
\label{separability}
Consider a binary of two point masses $m_1$ and $m_2$, where the
mass $m_1$ is endowed with a single axisymmetric current multipole
moment $S_l$ or axisymmetric mass multipole moment $I_l$. In this
section, we show that the Hamilton-Jacobi equation for this motion,
to linear order in the multipoles, to linear order in the mass ratio
and to the leading post-Newtonian order, is separable only for the
cases $S_1$ and $I_2$.

We choose the symmetry axis to be the $z$-axis and
 write the action for a general multipole as \bea S=\int dt && \left[\frac{1}{2}\left(\dot
r^2+r^2 \dot \theta^2 + r^2 \sin^2 \theta \dot
\varphi^2\right)+\frac{1}{r}\right.\nonumber\\
&& \left. ~ + f(r,\theta)+g(r,\theta)\dot \varphi
+E\right].\label{polars}\eea For mass moments, $g(r,\theta)=0$,
while for current moments $f(r,\theta)=0$. For an axisymmetric
multipole of order $l$, the functions $f$ and $g$ will be of the
form \be f(r,\theta)= \frac{c_l I_l P_l(\cos \theta)}{r^{l+1}},
{\;}{\;}g(r,\theta)=\frac{d_lS_l \sin\theta
\partial_\theta P_l(\cos \theta)}{r^l},\label{multifns}\ee where
$P_l(\cos \theta)$ are the Legendre polynomials and $c_l$ and $d_l$
are constants. We will work to linear order in $f$ and $g$. In Eq.
(\ref{polars}), we have added the energy term needed when doing a
change of time variables, cf. the discussion before Eq.
(\ref{action2}) in section \ref{sec2}. Since $\varphi$ is a cyclic
coordinate, $p_\varphi=L_z$ is a constant of motion and the system
has effectively only two degrees of freedom. Note that in the case
of a current moment, there will be correction term in $L_z$: \be
L_z=r^2\sin^2\theta \dot\varphi+g(r,\theta).\label{lzcurrent}\ee

Next, we switch to a different coordinate system $(\tilde r,
\tilde\theta, \varphi)$ defined by \bea r&=&\tilde
r + \alpha(\tilde r,\tilde \theta,L_z),\\
\theta&=& \tilde \theta + \beta (\tilde r,\tilde \theta,L_z), \eea
where the functions $\alpha$ and $\beta$ are yet undetermined. We
also define a new time variable $\tilde t$ by \be
dt=\left[1+\gamma(\tilde r,\tilde \theta, L_z)\right]d\tilde t. \ee
Since we work to linear order in $f$ and $g$, we can work to linear
order in $\alpha$, $\beta$, and $\gamma$. We then compute the action
in the new coordinates and drop the tildes. The Hamiltonian is given
by  \bea H &=& \frac{1}{2} p^2_r (1 + \gamma - 2 \alpha_{,r}) +
\frac{p_\theta^2}{
  2 r^2} (1 - \frac{2 \alpha}{r} - 2 \beta_{,\theta} + \gamma) \nonumber\\
  && +\frac{p_r p_\theta}{r^2} (-\alpha_{,\theta}-r^2
  \beta_{,r}) - E(1 + \gamma)\nonumber\\
   &&+ \frac{L_z^2}{ 2 r^2 \sin^2 \theta} (1 + \gamma - \frac{2 \alpha}{r }- 2 \beta \cot
  \theta)\nonumber\\
&&
 - \frac{1}{ r} (1 - \frac{\alpha}{r} + \gamma) - f-
 \frac{gL_z }{ r^2 \sin^2 \theta}\label{newhamiltonian} \eea
and the corresponding Hamilton-Jacobi equation is \bea 0&=& \left(
\frac{\partial W}{\partial r}\right)^2  \hat C_1+ \left(
\frac{\partial W}{\partial \theta }\right)^2\frac{\hat
C_2}{r^2}\nonumber\\
&& +2\left( \frac{\partial W}{\partial r}\right)\left(
\frac{\partial W}{\partial \theta}\right)\frac{\hat C_3}{r^2} +2\hat
V, \label{hj1} \eea where we have denoted \bea \hat
C_1&=&J(r,\theta)\left[1 + \gamma - 2 \alpha_{,r}\right]=1 + \gamma
- 2 \alpha_{,r}+j,\\
\hat C_2 &=& J(r,\theta)\left[1 - \frac{2 \alpha}{r} - 2
\beta_{,\theta} +
\gamma\right]\nonumber\\
&=&1 - \frac{2 \alpha}{r} - 2 \beta_{,\theta} + \gamma+j,\\
\hat C_3&=&J(r,\theta)\left[-\alpha_{,\theta}-r^2
\beta_{,r}\right]=-\alpha_{,\theta}-r^2 \beta_{,r},\\
\hat V&=& J(r,\theta)\left[\frac{L_z^2 }{2 r^2 \sin^2 \theta} (1 +
\gamma - \frac{2 \alpha}{r}- 2 \beta \cot
  \theta)\right.\nonumber\\
&&\left.{\;}{\;}{\;}{\;}{\;}{\;}{\;}{\;}{\;}{\;} - \frac{1}{ r} (1 -
\frac{ \alpha}{r} + \gamma) - E(1 + \gamma)\right.\nonumber\\
 &&\left.{\;}{\;}{\;}{\;}{\;}{\;}{\;}{\;}{\;}{\;} - f-
 \frac{gL_z}{r^2 \sin^2 \theta}\right]\nonumber\\
 &=&\frac{L_z^2 }{2 r^2 \sin^2 \theta} (1 +
\gamma - \frac{2 \alpha}{r}- 2 \beta \cot
  \theta+j)\nonumber\\
  &&- E(1 +
\gamma+j) - \frac{1}{ r} (1 - \frac{ \alpha}{r} + \gamma+j)
\nonumber\\
&& - f-
 \frac{gL_z}{r^2 \sin^2 \theta}.
 \eea
 The unperturbed problem is separable, so make the perturbed problem separable, we have multiplied the
 Hamilton-Jacobi equation by an arbitrary function $J(r,\theta)$, which can be expanded as
 $J(r,\theta)=1+j(r,\theta)$, where
$j(r,\theta)$ is a small perturbation.

To find a solution of the form $W=W_r(r)+W_\theta (\theta)$, we
first specialize to the case where $\hat C_3=0$: \be -\hat
C_3=\beta_{,r} r^2 + \alpha_{,\theta} =0. \label{nomixed} \ee We
differentiate Eq. (\ref{hj1}) with respect to $\theta$, using Eq.
(\ref{hj1}) to write $(dW_r/dr)^2$ in terms of
$(dW_\theta/d\theta)^2$ and then differentiate the result with
respect to $r$ to obtain \bea 0&=&\left( \frac{d W_\theta}{d
\theta}\right)^2\partial_r\left[\frac{\partial_\theta \hat C_2}{\hat
C_2}-\frac{\partial_\theta \hat C_1}{\hat
C_1}\right]\nonumber\\
&&+2\partial_r\left[r^2 \frac{\partial_\theta \hat V}{\hat
C_2}-\frac{r^2 \hat V\partial_\theta \hat C_1}{\hat C_1 \hat
C_2}\right].\label{hjdthetadr}\eea Expanding Eq. (\ref{hjdthetadr})
to linear order in the small quantities then yields the two
conditions for the kinetic and the potential part of the Hamiltonian
to be separable: \bea \label{egcond} 0&=&\partial_r\partial_\theta
\left( 2\alpha_{,r}-\frac{2 \alpha}{r}-2
\beta_{,\theta}\right),\\
0&=&\frac{L^2_z}{\sin^2\theta}\left(2\beta_{,r}
\cot^2\theta-3\beta_{,r \theta}\cot\theta +\beta_{,r}
\csc^2\theta\right)\nonumber\\
&&+\frac{L^2_z}{\sin^2\theta}\partial_r\left[-\frac{\alpha_{,\theta}}{r}
+\alpha_{,r\theta}\right]\nonumber\\
&&-\partial_r\partial_\theta\left[\frac{c_lI_l}{r^{l-1}} P_l(\cos
\theta)+\frac{d_lS_l L_z}{r^l\sin\theta}
\partial_\theta P_l(\cos \theta)\right]\nonumber\\
&&-\partial_r\left[r\left(2\alpha_{,r\theta}-\frac{\alpha_{,\theta}}{r}\right)
+2Er^2\alpha_{,r\theta}\right],\label{vcond} \eea where we have used
Eq. (\ref{multifns}) for $f$ and $g$. Therefore, the following
conditions must be satisfied: \bea M_4(\theta)-N(r) &=&
\frac{\alpha}{r}+\beta_{,\theta}-2 \alpha_{,r} ,\label{j4theta}\\
\label{j1theta} M_1(\theta)&=&2\beta \cot^2\theta+\beta
\csc^2\theta+\beta_{,\theta\theta}\nonumber\\
&&-3\beta_{,\theta}\cot\theta,\\\label{j2theta} M_2(\theta)&=&
r^2\partial_r(r^2\beta_{,r}),\\\label{j3theta} M_3(\theta)&=&
2r\alpha_{,r\theta}-\alpha_{,\theta}+\frac{I_l}{r^{l-1}}\partial_\theta
P_l(\cos \theta)\nonumber\\
&&-\frac{S_l L_z}{r^l}\partial_\theta (\csc\theta{\,}
\partial_\theta P_l(\cos \theta)).\eea Here, the functions $M$
and $N$ are arbitrary integration constants.

Solving the condition for the kinetic term to be separable, Eq.
(\ref{j4theta}), together with Eq. (\ref{nomixed}) gives the general
solution that goes to zero at large $r$ as \bea
\label{transformfs}\alpha &=& {A \over
r^{n-1}} \cos(n \theta + \nu),\\
\beta &=& -{A \over r^n} \sin(n\theta + \nu),\label{transformfns}
\eea where $A$ and $\nu$ are arbitrary and $n$ is an integer. These
functions must satisfy the conditions $(\ref{j1theta})$ --
$(\ref{j3theta})$ in order for the potential term to be separable as
well. To see when this will be the case, we start by considering Eq.
(\ref{j3theta}). Substituting the general ansatz $\alpha=a_1(r)
a_2(\theta)$ shows that $a_2'=P_l'$ or $a_2'=(csc \theta{\,} P_l')'$
depending on whether a mass or a current multipole is present. The
function $a_1(r)$ is
then determined from \be 0=2ra_1'-a_1+\begin{cases}c_lI_l/r^{(l-1)}\\
d_lS_lL_z/r^l
\end{cases}
\ee Hence, \be a_1=\begin{cases}[c_lI_l/(2l)] {\,}r^{(1-l)}\\
[d_lS_lL_z/(2l+1)]{\,}r^{-l}
\end{cases}
\ee so that we obtain for mass moments \be \alpha=\frac{c_lI_l}{2l}
\frac{P_l(\cos\theta)}{r^{l-1}}, {\;}{\;}{\;}{\;}{\;}{\;}
\beta=\frac{c_lI_l}{2l^2} \frac{P_l'(\cos\theta)}{r^{l}} \ee and for
current moments \bea
 \alpha &=&\frac{d_lS_lL_z}{2l+1} \frac{\csc
\theta P_l'(\cos\theta)}{r^{l}}, \\
 \beta &=&
\frac{d_lS_lL_z}{(2l+1)(l+1)} \frac{(\csc \theta{\,}
P_l'(\cos\theta))'}{r^{l+1}}, \eea where we have used the condition
(\ref{nomixed}) to solve for $\beta$.

Substituting this in Eq. (\ref{j2theta}) determines that $l=2$ for
mass moments and $l+1=2$ for current moments. For an $l=2$ mass
moment, conditions (\ref{j4theta}) and (\ref{j1theta}) are satisfied
as well, with $n=2$ and $\nu=0$. For the case of an $l=1$ current
moment, the extra term in $H$ is independent of $\theta$ anyway. But
for any other multipole interaction, the Hamilton-Jacobi equation
will not be separable. For example, for the current octupole
$S_{ijk}$, the last term in Eq. (\ref{newhamiltonian}) is
proportional to $ S_3 L_z (5\cos^2\theta-1)/r^5$ and is therefore
not separable. From Eq. (\ref{multifns}) one can see that, for a
general multipole, the functions $f$ or $g$ contain different powers
of $\cos\theta$ appearing with the same power of $r$ since the
Legendre polynomials can be expanded as \cite{arfken}: \be P_l(\cos
\theta)= \sum^{N}_{n=0} \frac{(-1)^n (2l-2n)!}{2^l n! (l-n)!
(l-2n)!}(\cos \theta)^{l-2n},\label{legendrep} \ee where $N=l/2$ for
even $l$ and $N=(l+1)/2$ for odd $l$. It will not be possible to
cancel all of these terms with (\ref{transformfs}) --
(\ref{transformfns}) for $l>2$.

The case when $\hat C_3$ is non-vanishing will only be separable if
all the coefficients are functions of $r$ or of $\theta$ only, and
if in addition, the potential also depends only on $r$ or on
$\theta$. Achieving this for our problem will not be possible
because the potential cannot be transformed to the form required for
separability.

\subsection{Derivation of non-existence of additional constants of the motion}
\label{poisson}
In this
subsection, we show using Poisson brackets that for a single
axisymmetric multipole interaction, to linear order in the multipole
and the mass ratio, a first integral analogous to the Carter
constant does not exist, except for the cases of mass quadrupole
and spin.

Suppose that such a constant does exist.  We write the Hamiltonian corresponding to the action (\ref{polars})
as $H=H_0+\delta H$ and the Carter-type constant as $K=K_0+\delta
K(p_r,p_\theta,L_z,r,\theta)$, where \bea
H_0&=&\frac{p_r^2}{2}+\frac{p_\theta^2}{2 r^2}+ \frac{L_z^2}{2 r^2
\sin^2\theta}-\frac{1}{r},\\
\delta H&=&-\frac{c_lI_l}{r^{l+1}}P_l(\cos\theta)-\frac{d_lS_l L_z}{
r^{l+2}\sin\theta}\partial_\theta P_l(\cos\theta),\\
K_0&=&p_\theta^2+\frac{L^2_z}{\sin^2\theta}.\eea
Computing the
Poisson bracket gives, to linear order in the perturbations
\bes
\bea
\label{condt:pde}
0&=&\{H_0,\delta K\}+\{\delta H,K_0\} \\
&=&\frac{d}{dt}\delta K+\{\delta
H,K_0\},
\eea
\label{fisch1}
\ees
where we have used that $\{H_0, K_0\}=0$
and the fact that $\{H_0, \delta K\}=d(\delta K)/dt$. Here, $d/dt$
denotes the total time derivative along an orbit $(r(t), \theta (t),
p_r (t), p_\theta (t))$ of $H_0$ in phase space.
 The partial differential
equation (\ref{condt:pde}) for $\delta K$ thus reduces to a set of
ordinary differential equations
that can be integrated along the individual orbits in phase space.

The unperturbed motion for a bound orbit is in a plane, so we can
switch from spherical to plane polar coordinates $(r,\psi)$. In
terms of these coordinates, we have $H_0=p_r^2/2+p_\psi^2/2,{\,} $
$K_0=p^2_\psi$, and $\cos\theta=\sin \iota \sin(\psi +\psi_0),$ with
$\cos \iota=L_z/\sqrt{K}$ and the constant $\psi_0$ denoting the
angle between the direction of the periastron and the intersection
between the orbital and equatorial plane. Then Eq. (\ref{fisch1})
becomes \bea \label{dtK} \frac{d}{dt}\delta
K&=&\eta(t),\\
\eta(t)&=&-\frac{2  p_\psi ~ d_l  S_l L_z}{\sin \iota ~
r^{l+2}(t)}\partial_\psi \left(\frac{\partial_\psi P_l(\sin \iota
\sin(\psi(t)+\psi_0))}{\cos(\psi(t)+\psi_0)}\right)\nonumber\\
&&+\frac{2p_\psi ~ c_lI_l}{r^{l+1}(t)}\partial_\psi P_l(\sin \iota
\sin(\psi(t)+\psi_0)) .\label{fisch2}\eea

For unbound orbits, one can always integrate Eq. (\ref{dtK}) to
determine $\delta K$. However, for bound periodic orbits there is a
possible obstruction: the solution for the conserved quantity
$K_0+\delta K$ will be single valued if and only if the integral of
the source over the closed orbit vanishes, \be \oint^{T_{\rm orb}}_0
\eta(t) dt=0. \label{singleval}\ee Here, $T_{\rm orb}$ is the
orbital period. In other words, the partial differential equation
(\ref{fisch1}) has a solution $\delta K$ if and only if the
condition (\ref{singleval}) is satisfied. This is the same condition
as obtained by the Poincare-Mel'nikov-Arnold method, a technique for
showing the non-integrability and existence of chaos in certain
classes of perturbed dynamical systems \cite{Melnikov}.

Thus, it suffices to show that the condition (\ref{singleval}) is
violated for all multipoles other than the
spin and mass quadrupole.
To perform the integral in Eq.\ (\ref{singleval}), we use the
parameterization for the unperturbed motion, $r=K/(1+e\cos\psi)$ and
$dt/d\psi=K^{3/2}/(1+e\cos\psi)^2$, so that the condition for the
existence of a conserved quantity $K_0+\delta K$ becomes
\begin{widetext}
\be \int^{2\pi}_0 d\psi \left[c_lI_l
(1+e\cos\psi)^{l-1}\partial_\psi
P_l(\sin\iota\sin(\psi+\psi_0))-\frac{d_lS_l
L_z}{K\sin\iota}(1+e\cos\psi)^l\partial_\psi\left(\frac{\partial_\psi
P_l(\sin\iota\sin(\psi+\psi_0))} {\cos(\psi+\psi_0)}
\right)\right]=0 .\label{existsoln} \ee
 In terms of the variable
$\chi=\psi+\psi_0-\pi/2$, Eq. (\ref{existsoln}) can be written as
\bea 0&=&\int^{2\pi}_0 d\chi
c_lI_l\left[1+e(\sin\psi_0\cos\chi-\cos\psi_0\sin\chi)\right]^{l-1}\frac{d}{d\chi}P_l(\sin\iota\cos\chi)\nonumber\\
&+&\int^{2\pi}_0 d\chi\frac{d_lS_l
L_z}{\sin\iota}\left[1+e(\sin\psi_0\cos\chi-\cos\psi_0\sin\chi)\right]^{l}
\frac{d}{d\chi}\left(\frac{1}{\sin\chi}\frac{d}{d\chi}P_l(\sin\iota\cos\chi)\right).
\label{fischi} \eea Inserting the expansion (\ref{legendrep}) for
$P_l(\cos\chi)$, taking the derivatives, and using the binomial
expansion for the first term in Eq. (\ref{fischi}), we get \bea
0&=&c_lI_l\sum^N_{n=0}\sum^{l-1}_{j=0}A_{lnjk}{\,}e^j(\sin\iota)^{l-2n}(\sin\psi_0)^k(\cos\psi_0)^{j-k}
\int^{2\pi}_0d\chi {\,} (\sin\chi)^{j-k+1}
(\cos\chi)^{k+l-2n-1}\nonumber\\
&+&\frac{d_lS_lL_z}{K}\sum^N_{n=0}\sum^{l}_{j=0}B_{lnjk}{\,}e^j(\sin\iota)^{l-2n-1}
(\sin\psi_0)^k(\cos\psi_0)^{j-k}\int^{2\pi}_0d\chi {\,}
(\sin\chi)^{j-k+1} (\cos\chi)^{k+l-2n-2}.\label{fischintegrals} \eea
The coefficients $A_{lnkj}$ and $B_{lnkj}$ are \be
A_{lnkj}=\frac{(-1)^{n+k+1}(l-1)!(2l-2n)!}{2^ln!(l-1-j)!k!(j-k)!(l-n)!(l-2n-1)!},{\;}
{\;}{\;}
B_{lnkj}=\frac{(-1)^{n+k}l!(2l-2n)!}{2^ln!(l-j)!k!(j-k)!(l-n)!(l-2n-2)!}.
\ee  The only non-vanishing contribution to the integrals in Eq.
(\ref{fischintegrals}) will come from terms with even powers of both
$\cos\chi$ and $\sin\chi$. These can be evaluated as multiples of
the beta function: \bea
0&=&c_lI_l\sum^N_{n=0}\sum^{l-1}_{j=0}C_{lnjk}{\,}e^j(\sin\iota)^{l-2n}(\sin\psi_0)^k(\cos\psi_0)^{j-k}
{\;}\delta_{(j-k+1),{\rm even}}{\,}\delta_{(l+k-1),{\rm even}}\nonumber\\
&+&\frac{d_lS_lL_z}{K}\sum^N_{n=0}\sum^{l}_{j=0}D_{lnjk}{\,}e^j(\sin\iota)^{l-2n-1}
(\sin\psi_0)^k(\cos\psi_0)^{j-k}{\;}\delta_{(j-k+1),{\rm
even}}{\,}\delta_{(l+k),{\rm even}}. \label{fischresults} \eea Here,
the coefficients are \be
C_{lnjk}=\frac{2\Gamma(\frac{j}{2}-\frac{k}{2}+1)\Gamma(\frac{k}{2}+\frac{l}{2}-n)}
{\Gamma(\frac{j}{2}+\frac{l}{2}-n+1)}A_{lnkj},{\;}{\;}{\;}
D_{lnjk}=\frac{2\Gamma(\frac{j}{2}-\frac{k}{2}+1)\Gamma(\frac{k}{2}+\frac{l}{2}-n-\frac{1}{2})}
{\Gamma(\frac{j}{2}+\frac{l}{2}-n+\frac{3}{2})}B_{lnkj}\ee
\end{widetext}
Eq. (\ref{fischresults}) shows that for even $l$, terms with
$j=$even (odd) and $k=$odd (even) give a non-vanishing contribution
for the case of a mass (current) multipole, and hence $K_0+\delta K$
is not a conserved quantity for the perturbed motion. Note that
terms with $j=$even and $k=$odd for even $l$ occur only for $l>3$,
so for $l=2$ the mass quadrupole term in Eq. (\ref{fischresults})
vanishes and therefore there exists an analog of the Carter
constant, which is consistent with our results of Sec. {\ref{sec1}}
and our separability analysis. For odd $l$, terms with $j=$odd
(even) and $k=$even (odd) are finite for $I_l$ ($S_l$). Note that
for the case $l=1$ of the spin, the derivatives with respect to
$\chi$ in Eq. (\ref{fischi}) evaluate to zero, so in this case there
also exists a Carter-type constant. These results show that for a
general multipole other than $I_2$ and $S_1$, there will not be a
Carter-type constant for such a system.

\subsubsection{Exact vacuum spacetimes}

Our result on the non-existence of a Carter-type constant can be
extended, with mild smoothness assumptions, to falsify the conjecture that
all exact, axisymmetric vacuum spacetimes posess a third constant
of the motion for geodesic motion.
Specifically, we fix a multipole order $l$, and we assume:
\begin{itemize}
\item There exists a one parameter family
$$
\left(M,g_{ab}(\lambda)\right)
$$
of spacetimes, which is smooth in the parameter $\lambda$, such that $\lambda=0$ is Schwarzschild,
and each spacetime $g_{ab}(\lambda)$ is stationary and axisymmetric
with commuting Killing fields $\partial/\partial t$ and
$\partial
/ \partial \phi$, and such that all the mass and current multipole
moments of the spacetime vanish except for the one of order $l$.
On physical grounds, one expects a one parameter family of metrics
with these properties to exist.

\item We denote by $H(\lambda)$ the Hamiltonian on the tangent bundle
  over $M$ for geodesic motion in the metric $g_{ab}(\lambda)$.
By hypothesis, there exists for each $\lambda$
a conserved quantity $M(\lambda)$ which is functionally independent of
the conserved energy and angular momentum.  Our second assumption is
that $M(\lambda)$ is differentiable in $\lambda$ at $\lambda=0$.
One would expect this to be true on physical grounds.

\item We assume that the conserved quantity $M(\lambda)$ is invariant
  under the symmetries of the system:
$$
{\cal L}_{\vec \xi} M(\lambda) = {\cal L}_{\vec \eta} M(\lambda) =0,
$$
where ${\vec \xi}$ and ${\vec \eta}$ are the natural extensions to the
8 dimensional phase space of the Killing vectors $\partial/\partial t$
and $\partial / \partial \phi$.  This is a very natural assumption.

\end{itemize}

These assumptions, when combined with our result of the previous
section, lead to a contradiction, showing that the conjecture is
false under our assumptions.

To prove this, we start by noting that
$M(0)$ is a conserved quantity for geodesic motion in
Schwarzschild, so it must be possible to express it as some function $f$
of the three independent conserved quantities:
\be
M(0) = f(E,L_z,K_0).
\ee
Here $E$ is the energy, $L_z$ is the angular
momentum, and $K_0$ is the Carter constant.
Differentiating the exact relation $\left\{ H(\lambda), M(\lambda) \right\}=0$ and
evaluating at $\lambda=0$ gives
\be
\{ H_0, M_1 \} =
{\partial f \over
\partial E} \{ E, H_1\} + {\partial f \over \partial L_z} \{ L_z,
H_1 \} + {\partial f \over \partial K_0} \{ K_0, H_1 \},
\ee
where $H_0 = H(0)$, $H_1 = H'(0)$, and $M_1 = M'(0)$.
As before, we can regard this is a partial differential equation that
determines $M_1$, and a necessary condition for solutions to exist and be
single valued is that the integral of the right hand side over any
closed orbit must vanish:
\be \oint \left[
{\partial f \over
\partial E} \{ E, H_1\} + {\partial f \over \partial L_z} \{ L_z,
H_1 \} + {\partial f \over \partial K_0} \{ K_0, H_1 \}\right]=0.
\label{condt:pde1}\ee
Now strictly speaking, there are no closed orbits
in the eight dimensional phase space.  However, the argument of the
previous section applies to orbits which are closed in the four dimensional space
with coordinates $(r,\theta,p_r,p_\theta)$,
since by the third assumption above everything is
independent of $t$ and $\phi$, and $p_t$ and $p_\phi$ are conserved.  Here
$(t,r,\theta,\phi)$ are Schwarzschild coordinates  and
$(p_t,p_r,p_\theta,p_\phi)$ are the corresponding conjugate momenta.

Next, we can pull the partial derivatives ${\partial f / \partial E}$
etc. outside of the integral.  It is then easy to see that the first
two terms vanish, since there do exist a conserved energy and a
conserved $z$-component of angular
momentum for the perturbed system.
Thus, Eq. (\ref{condt:pde1})
reduces to
\be {\partial f \over \partial K_0} \oint \{ K_0,H_1
\}=0.
\label{condt:pde2}
\ee
Since $M(0)$ is functionally independent of $E$ and $L_z$, the
prefactor $\partial f / \partial K_0$ must be nonzero, so we obtain
\be
\oint \{ K_0,H_1
\}=0.
\label{condt:pde3}
\ee

The result (\ref{condt:pde3}) applies to fully relativistic orbits in
Schwarzschild.  We need to take the Newtonian limit of this result in
order to use the result we derived in the previous section.
However, the Newtonian limit is a little subtle since Newtonian orbits are
closed and generic relativistic orbits are not closed.  We now discuss
how the limit is taken.

The integral (\ref{condt:pde3}) is taken over any closed orbit in the four dimensional
phase space $(r,\theta,p_r,p_\theta)$ which corresponds to a geodesic in Schwarzschild.
Such orbits are non generic; they are the orbits for which the ratio
between the radial and angular frequencies $\omega_r$ and
$\omega_\theta$ is a rational number.
We denote by $q_r$ and $q_\theta$ the angle variables corresponding to
the $r$ and $\theta$ motions \cite{Schmidt}.
These variables evolve with proper time $\tau$ according to
\bes
\label{qse}
\bea
q_r &=& q_{r,0} + \omega_r \tau, \\
q_\theta &=& q_{\theta,0} + \omega_\theta \tau,
\eea
\ees
where $q_{r,0}$ and $q_{\theta,0}$ are the initial values. We denote
the integrand in Eq.\ (\ref{condt:pde3}) by
$$
{\cal I}(q_r,q_\theta,a,\varepsilon,\iota),
$$
where ${\cal I}$ is some function, and $a$, $\varepsilon$ and $\iota$
are the parameters of the geodesic defined by Hughes \cite{hughes}
(functions of $E$, $L_z$ and $K_0$).
The result (\ref{condt:pde3}) can be written as
\be
\frac{1}{T} \int_{-T/2}^{T/2} d\tau \, {\cal
  I}[q_r(\tau),q_\theta(\tau),a,\varepsilon,\iota] =0,
\label{condt44}
\ee
where $T = T(a,\varepsilon,\iota)$ is the period of the $r$, $\theta$
motion.

Since the variables $q_r$ and $q_\theta$ are periodic with period $2
\pi$, we can express the function ${\cal I}$ as a Fourier series
\be
{\cal I}(q_r,q_\theta,a,\varepsilon,\iota) = \sum_{n,m=-\infty}^\infty
{\cal I}_{nm}(a,\varepsilon,\iota) e^{i n q_r + i m q_\theta}.
\label{fseries}
\ee
Now combining Eqs.\ (\ref{qse}), (\ref{condt44}) and (\ref{fseries}) gives
\bea
0 &=& \sum_{n,m=-\infty}^\infty
{\cal I}_{nm}(a,\varepsilon,\iota) e^{i n q_{r,0} + i m q_{\theta,0}}
\nonumber \\
&& \times {\rm Si} \left[ (n \omega_r + m \omega_\theta) T/2 \right],
\eea
where ${\rm Si}(x) = \sin(x)/x$.  Since the initial conditions
$q_{r,0}$ and $q_{\theta,0}$ are arbitrary, it follows that
\be
{\cal I}_{nm}(a,\varepsilon,\iota)
 {\rm Si} \left[ (n \omega_r + m \omega_\theta) T/2 \right] =0
\label{aaa}
\ee
for all $n$, $m$.

Next, for closed orbits the ratio of the frequencies must be a rational
number, so
\be
\frac{w_r}{w_\theta} = \frac{p}{q},
\label{rational}
\ee
where $p$ and $q$ are integers with no factor in common.
These integers depend on $a$, $\varepsilon$ and $\iota$.  The period
$T$ is given by $2 \pi/T = q \omega_r = p \omega_\theta$. The second
factor in Eq.\ (\ref{aaa}) now simplifies to
\be
{\rm Si} \left[ \frac{ (np + mq) \pi} {p q} \right],
\ee
which vanishes if and only if
\be
n = {\bar n} q, \ \ \ m = {\bar m} p, \ \ \ {\bar n} + {\bar m} \ne 0,
\label{condt4}
\ee
for integers ${\bar n}$, ${\bar m}$.  It follows that
\be
{\cal I}_{nm}(a,\varepsilon,\iota) =0
\label{anss1}
\ee
for all $n$, $m$ except for values of $n$, $m$ which satisfy the
condition (\ref{condt4})

Consider now the Newtonian limit, which is the limit $a \to \infty$
while keeping fixed $\varepsilon$ and $\iota$ and the mass of the
black hole.  We denote by ${\cal I}_{{\rm N}}(q_r,q_\theta,a,\varepsilon,\iota)$
the Newtonian limit of the function ${\cal
I}(q_r,q_\theta,a,\varepsilon,\iota)$.
The integral (\ref{condt44}) in the Newtonian limit is given by the above
computation with $p=q=1$, since $\omega_r = \omega_\theta$ in this
limit.  This gives
\be
\frac{1}{T} \oint d\tau {\cal I}_{\rm N} = \sum_{n=-\infty}^\infty
{\cal I}_{{\rm N}\,n,-n}(a,\varepsilon,\iota) \, e^{i n (q_{r,0} - q_{\theta,0})},
\ee
where ${\cal I}_{{\rm N}\,nm}$ are the Fourier components of ${\cal
  I}_{\rm N}$.
In the previous subsection, we showed that this function is non-zero,
which implies that there exists a value $k$ of $n$ for which
${\cal I}_{{\rm N}\,k,-k} \ne 0$.

Now as $a \to \infty$, we have $\omega_r / \omega_\theta \to 1$, and hence
from Eq.\ (\ref{rational}) there exists a critical value $a_c$ of $a$ such that
the values of $p$ and $q$ exceed $k$ for all closed orbits with $a > a_c$.
(We are keeping fixed the values of $\varepsilon$ and $\iota$).
It follows from Eqs.\ (\ref{condt4}) and (\ref{anss1}) that
\be
\frac{ {\cal I}_{k,-k}(a,\varepsilon,\iota)}{
{\cal I}_{{\rm N}\,k,-k}(a,\varepsilon,\iota)}=0
\ee
for all such values of $a$.  However this contradicts the fact that
\be
\frac{ {\cal I}_{k,-k}(a,\varepsilon,\iota)}{
{\cal I}_{{\rm N}\,k,-k}(a,\varepsilon,\iota)} \to 1
\ee
as $a \to \infty$.
This completes the proof.

Hence, if the three assumptions listed at the start of this
subsection are satisfied, then the conjecture that all vacuum,
axisymmetric spacetimes possess a third constant of the motion is
false.

Finally, it is sometimes claimed in the classical dynamics
literature that perturbation theory is not a sufficiently powerful
tool to assess whether the integrability of a system is preserved
under deformations. An example that is often quoted is the Toda
lattice Hamiltonian \cite{Birol,Yoshida}. This system is integrable
and admits a full set of constants of motion in involution. However,
if one approximates the Hamiltonian by Taylor expanding the
potential about the origin to third order, one obtains a system
which is not integrable. This would seem to indicate that
perturbation theory can indicate a non-integrability, while the
exact system is still integrable.

In fact, the Toda lattice example does not invalidate the method of
proof we use here. If we write the Toda lattice Hamiltonian as
$H({\bf q}, {\bf p})$, then the situation is that $H( \lambda {\bf
q}, {\bf p})$ is integrable for $\lambda=1$, but it is not
integrable for $0 < \lambda < 1$.  Expanding $H(\lambda {\bf q},
{\bf p})$ to third order in $\lambda$ gives a non-integrable
Hamiltonian.  Thus, the perturbative result is not in disagreement
with the exact result for $0 < \lambda < 1$, it only disagrees with
the exact result for $\lambda=1$.  In other words, the example shows
that perturbation theory can fail to yield the correct result for
finite values of $\lambda$, but there is no indication that it fails
in arbitrarily small neighborhoods of $\lambda=0$.  Our application
is qualitatively different from the Toda lattice example since we
have a one parameter family of Hamiltonians $H(\lambda)$ which by
assumption are integrable for all values of $\lambda$.

\section{conclusion}
We have examined the effect of an axisymmetric quadrupole moment $Q$
of a central body on test particle inspirals, to linear order in
$Q$, to the leading post-Newtonian order, and to linear order in the
mass ratio. Our analysis shows that a natural generalization of the
Carter constant can be defined for the quadrupole interaction. We
have also analyzed the leading order spin self-interaction effect
due to the scattering of the radiation off the spacetime curvature
due to the spin. Combining the effects of the quadrupole and the
leading order effects linear and quadratic in the spin, we have
obtained expressions for the instantaneous as well as time-averaged
evolution of the constants of motion for generic orbits under
gravitational radiation reaction, complete at $O(a^2\epsilon^4)$. We
have also shown that for a single multipole interaction other than
$Q$ or spin, in our approximations, a Carter-type constant does not
exist. With mild additional assumptions, this result can be extended
to exact spacetimes and falsifies the conjecture that
all axisymmetric vacuum spacetimes possess a third
constant of motion for geodesic motion.

\section{Acknowledgments}
This research was partially supported by NSF grant PHY-0457200.
We thank Jeandrew Brink for useful correspondence.

\appendix
\section{Time variation of quadrupole: order of magnitude estimates}
In this appendix, we give an estimate of the timescale $T_{\rm
evol}$ for the quadrupole to change. The analysis in the body of
this paper is valid only when $T_{\rm evol}\gg T_{\rm rr}$, where
$T_{rr}$ is the radiation reaction time, since we have neglected the
time evolution of the quadrupole. We distinguish between two cases:
(i) when the central body is exactly nonspinning but has a
quadrupole, and (ii) when the central body has finite spin in
addition to the quadrupole.

\subsection{Estimate of the scaling for the nonspinning case} For
the purpose of a crude estimate, the relevant interaction is the
tidal interaction with energy \be Q_{ij}{\cal{E}}_{ij}\sim
-\frac{m_2}{r^3}\bar Q I \cos^2\theta, \ee where ${\cal{E}}_{ij}$ is
the tidal field, $\theta$ is the angle between the symmetry axis and
the normal to the orbital plane of $m_2$, and we have written the
quadrupole as $Q\sim \bar Q I$, where $\bar Q$ is dimensionless and
$I$ is the moment of inertia.
For small deviations from equilibrium, the relevant piece of the
Lagrangian is schematically  \be L\sim I\dot\psi^2+\bar Q I
\frac{m_2}{r^3}\psi^2.\ee We define the evolution timescale $T_{\rm
evol}$ to be the time it takes for the angle to change by an amount
of order unity, and since the amplitude of the oscillation scales
roughly as $\sim m_2/m_1$, the evolution time scales as \be
T^{-2}_{\rm evol}\sim \frac{m^2_2}{m^2_1}\bar Q
\left(\frac{m_2}{M}\right)\omega^2_{\rm
orbit},\label{nospintprec}\ee  where $\omega^2_{\rm orbit}=M/r^3.$
Thus, the ratio of the evolution timescale compared to the radiation
reaction timescale scales as \be T_{\rm evol}/T_{\rm rr}\sim
\left(1/\sqrt{\bar
Q}\right)\frac{m_1}{m_2}\left(\frac{\mu}{M}\right)^{1/2}\left(\frac{M}{r}\right)^{5/2}.
\ee

\subsection{Estimate of the scaling for the spinning case} When the
body is spinning the effect of the tidal coupling is to cause a
precession. For the purpose of this estimate, we calculate the
torque on $m_1$ due to the companion's Newtonian field. The torque
${\bf N}$ scales as \be N_i\sim \epsilon_{imj}Q_{mk}{\cal E}_{jk}.
\ee We assume that the precession is slow, i.e. \be \omega_{\rm
prec}\ll \bar S/m_1 \left(\frac{m_2}{M}\right),\label{slowprec}\ee
where $\omega_{\rm prec}$ is the precession frequency and $\bar
S=S/m_1^2$ is the dimensionless spin. This gives the approximate
scaling of the precession timescale as (cf. \cite{Goldstein})\be
T_{\rm prec}/T_{\rm rr}\sim \frac{\bar S }{\bar
Q}\left(\frac{M}{r}\right). \label{spintprec}\ee and the evolution
timescale is thus \be T_{\rm evol}/T_{\rm rr}\sim
\frac{m_1}{m_2}\frac{\bar S }{\bar
Q}\left(\frac{M}{r}\right).\label{spintevol} \ee

Because of our assumption (\ref{slowprec}) that the precession is
slow, equation (\ref{spintevol}) is valid only when \be 1 \gg
\left(\frac{\mu}{M}\right) \frac{\bar S ^2}{\bar Q}
\left(\frac{r}{M}\right)^3.\label{precvalid}\ee

When $\bar S$ is sufficiently small that the condition
(\ref{precvalid}) is violated, the relevant timescale is instead
given by Eq. (\ref{nospintprec}).

\subsection{Application to Kerr inspirals} For Kerr inspirals, \be
\bar S\sim a,{\;}{\;}{\;}\bar Q\sim a^2,{\;}{\;}{\;}\mu/M\ll
1{\;}{\rm and}{\;}{\;}r\sim M.\ee Therefore, the condition
(\ref{precvalid}) is satisfied, and the precession time is longer
than the radiation reaction time by \be T_{\rm prec}/T_{\rm rr}\sim
\frac{1}{a} {\,} \left(\frac{M}{r}\right).\ee  Note that for Kerr
inspirals, since $r\sim M$ both formulas (\ref{nospintprec}) and
(\ref{spintprec}) give the same scaling.

Moreover, for Kerr inspirals, the amplitude of the precession will
be small, of order the mass ratio $\mu/M$. This is because of
angular momentum conservation: in the relativistic regime, the
orbital angular momentum is a factor of $\mu/M$ smaller than the
angular momentum of the black hole and can therefore not cause a
large precession amplitude. Even if the orbital angular momentum at
infinity is large, most of it will be radiated away as outgoing
gravitational waves during the earlier phase of the inspiral. This
factor of $\mu/M$ is taken into account when we consider the
evolution timescale, which for Kerr inspirals reduces to \be T_{\rm
evol}/T_{\rm rr}\sim \left(\frac{M}{\mu}\right)
\left(\frac{1}{a}\right)\left(\frac{M}{r}\right).\ee Since $1/a\geq
1$, $M/r\sim 1$ and $M/\mu\ll 1$, the evolution time is long
compared to the radiation reaction time and we can neglect the time
variation of the quadrupole at leading order.

\section{Computation of time averaged fluxes}
\subsection{Averaging method that parallels fully relativistic
averaging}
 We start by noting that the differential
equations (\ref{s1}) and (\ref{s2}) governing the $\tilde r$ and
$\tilde \theta$ motions decouple if we define a new time parameter
$\hat t$ by \be d\hat t=\frac{1}{\tilde r^2}d\tilde
t.\label{minotime}\ee This is the analog of the Mino time parameter
for geodesic motion in Kerr \cite{Mino2003}. The equations of motion
(\ref{s1})--(\ref{s3}) then become \bea \left(\frac{d\tilde r}{d\hat
t}\right)^2&=& \hat V_{\tilde r}(\tilde r),\label{s1mino}\\
 \hat V_{\tilde
r}(\tilde r)&=&2E{\tilde r}^4+2{\tilde r}^3-K{\tilde
r}^2-4SL_z{\tilde r}\nonumber\\
&&+\frac{Q}{2}\left(\tilde r-2 L^2_z\right),\\
 \left(\frac{d\tilde \theta}{d\hat t}\right)^2&=& \hat
V_{\tilde \theta}(\tilde \theta),\label{s2mino}\\
 \hat V_{\tilde \theta}(\tilde
\theta)&=&K-\frac{L^2_z}{\sin^2\tilde\theta}
-QE\cos 2\tilde \theta ,\\
 \left(\frac{d\varphi}{d\hat t}\right)&=& \hat
V_{\varphi \tilde r}(\tilde r)+\hat V_{\varphi\tilde\theta}(\tilde
\theta),\label{s3mino}\\
 \hat V_{\varphi \tilde r}(\tilde r)&=&\frac{QL_z}{\tilde
r^2},{\;}{\;}{\;}{\;}{\;}{\;} \hat V_{\varphi \tilde \theta}(\tilde
\theta)=\frac{L_z}{\sin^2\tilde\theta}. \eea The parameters $t$ and
$\hat t$ are related by: \bea \frac{dt}{d\hat t}&=&\hat V_{t\tilde
r}(\tilde r)+\hat V_{t\tilde\theta}(\tilde \theta)\label{dtmino}\\
\hat V_{t\tilde r}(\tilde r)&=&{\tilde r}^2,{\;}{\;}{\;}{\;}{\;}{\;}
\hat V_{t\tilde\theta}(\tilde\theta)=\frac{Q}{2}\cos 2\tilde \theta
. \eea It follows from Eqs. (\ref{s1mino}) and (\ref{s2mino}) that
the functions $\tilde r(\hat t)$ and $\tilde \theta(\hat t)$ are
periodic; and we denote their periods by $\Lambda_{\tilde r}$ and
$\Lambda_{\tilde \theta}$. We define the fiducial motion associated
with the constants of motion $E$, $L_z$ and $K$ to be the motion
with the initial conditions $\tilde r(0)=\tilde r_{\rm min}$ and
$\tilde \theta(0)=\tilde \theta_{\rm min}$, where $\tilde r_{\rm
min}$ and $\tilde \theta_{\rm min}$ are given by the vanishing of
the right-hand sides of Eqs. (\ref{s1mino}) and (\ref{s2mino})
respectively. The functions $\hat r(\hat t)$ and $\hat \theta(\hat
t)$ associated with this fiducial motion are given by \bea
\int^{\hat r(\hat t)}_{\tilde r_{\rm min}}\frac{d\tilde r}{\pm
\sqrt{\hat V_{\tilde r}(\tilde r)}}
&=&\hat t,\label{fidr}\\
\int^{\hat \theta(\hat t)}_{\tilde \theta_{\rm min}}\frac{d\tilde
\theta}{\pm \sqrt{\hat V_{\tilde \theta}(\tilde \theta)}}&=&\hat t .
\label{fidtheta} \eea From Eq. (\ref{dtmino}) it follows that \be
t(\hat t)=t_0+\int^{\hat t}_0 dt'\left(\hat V_{t\tilde r}[\tilde
r(t')]+\hat V_{t\tilde \theta}[\tilde \theta(t')]\right),
\label{tofthat}\ee where $t_0=t(0)$. Next, we define the constant
$\Gamma$ to be the following average value: \be \Gamma
=\frac{1}{\Lambda_{\tilde r}}\int^{\Lambda_{\tilde r}}_0dt'\hat
V_{t{\tilde r}}[\hat r(t')]+ \frac{1}{\Lambda_{\tilde
\theta}}\int^{\Lambda_{\tilde \theta}}_0dt'\hat V_{t{\tilde
\theta}}[\hat \theta(t')]. \ee Then we can write $t(\hat t)$ as a
sum of a linear term and terms that are periodic: \be t(
t)=t_0+\Gamma \hat t+\delta t(\hat t),\ee where $\delta t(\hat t)$
denotes the oscillatory terms in Eq. (\ref{tofthat}).

To average a function over the time parameter $\hat t$, it is
convenient to parameterize $\tilde r$ and $\tilde \theta$ in terms
of angular variables as follows. For the average over $\tilde
\theta$ we introduce the parameter $\chi$ by \be \cos^2\hat
\theta(\hat t)=z_-\cos^2\chi, \label{chidef}\ee where $z_-=\cos^2
\tilde \theta_-$ with $z_-$ being the smaller root of Eq.
(\ref{s2mino}): \be z_\pm=\frac{1}{2\beta}\left[K+3 QE\pm
\sqrt{\left(K-QE\right)^2+4QEL^2_z}\right] \label{zpm} \ee and where
$\beta=2QE$. Then from the definition (\ref{fidtheta}) of $\hat
\theta$ together with Eq. (\ref{s2mino}) and the requirement that
$\chi$ increases monotonically with $\hat t$ we obtain \be
\frac{d\chi}{d\hat t}=\sqrt{\beta\left(z_+-z_-\cos^2\chi\right)}.
\label{dchidt} \ee Then we can write the average over $\hat t$ of a
function $F_{\tilde \theta}(\hat t)$ which is periodic with period
$\Lambda_{\tilde \theta}$ in terms of $\chi$ as \bea \langle
F_{\tilde \theta}\rangle_{\hat t}&=&\frac{1}{\Lambda_{\tilde
\theta}}\int^{\Lambda_{\tilde \theta}}_0 d\hat t F_{\tilde \theta}
(\hat t)\nonumber\\
&=&\frac{1}{\Lambda_{\tilde
\theta}}\int^{2\pi}_0d\chi\frac{F_{\tilde \theta}[\hat
t(\chi)]}{\sqrt{\beta\left(z_+-z_-\cos^2\chi\right)}},
\label{thetaav} \eea where \be \Lambda_{\tilde \theta}=\int^{2\pi}_0
d\chi\frac{1}{\sqrt{\beta\left(z_+-z_-\cos^2\chi\right)}}.\label{Lambdatheta}
\ee Similarly, to average a function $F_{\tilde r}(\hat t)$ that is
periodic with period $\Lambda_{\tilde r}$, we introduce a parameter
$\xi$ via \be \tilde r=\frac{p}{1+e \cos \xi},\label{rxi}\ee where
the parameter $\xi$ varies from $0$ to $2\pi$ as $\tilde r$ goes
through a complete cycle. Then, \bea \frac{d\xi}{d\hat
t}&=&P(\xi), \label{dxidthat}\\
P(\xi)&\equiv& \left(\hat V_{\tilde r}[\tilde
r(\xi)]\right)^{1/2}\left[\frac{p e\mid \sin \xi \mid}{\left(1+e\cos
\xi\right)^2}\right]^{-1} \eea The average over $\hat t$ of
$F_{\tilde r}(\hat t)$ can then be computed from \be \langle
F_{\tilde r}\rangle_{\hat t}=\frac{\int^{2\pi}_0 d\xi {\,}F_{\tilde
r}/P(\xi)}{\int^{2\pi}_0 d\xi /P(\xi)}. \label{rav} \ee Now, a
generic function $F_{\tilde r,\tilde \theta}[\tilde r(\hat t),\tilde
\theta (\hat t)]$ will be biperiodic in $\hat t$: $F_{\tilde
r,\tilde \theta}[\tilde r(\hat t+\Lambda_{\tilde r}),\tilde \theta
(\hat t+\Lambda_{\tilde \theta})]=F_{\tilde r,\tilde \theta}[\tilde
r(\hat t),\tilde \theta (\hat t)]$. Combining the results
(\ref{thetaav}) and (\ref{rav}) we can write its average as a double
integral over $\chi$ and $\xi$ as \be \langle F_{\tilde r,\tilde
\theta}\rangle_{\hat
t}=\frac{1}{\Lambda_{\tilde\theta}\Lambda_{\tilde
r}}\int^{2\pi}_0d\chi \int^{2\pi}_0 d\xi \frac{F_{\tilde r,\tilde
\theta}[\tilde r(\xi),\tilde
\theta(\chi)]}{\sqrt{\beta\left(z_+-z_-\cos^2\chi\right)}P(\xi)}.
\label{average} \ee

To compute the time average of $\dot E$, $\dot L_z$, and $\dot K$,
we need to convert the average of a function over $\hat t$
calculated from (\ref{average}) to the average over $t$. As
explained in detail in \cite{scalar}, in the adiabatic limit we
can choose a time interval $\Delta t$ which is long compared to
the orbital timescale but short compared to the radiation reaction
time. From Eq. (\ref{tofthat}) we have $\Delta t=\Gamma \hat
t+{\rm osc. terms}$. The oscillatory terms will be bounded and
will therefore be negligible in the adiabatic limit, so we have to
a good approximation \be \langle \dot E
\rangle_t=\frac{1}{\Gamma}\langle \dot E{\;}\hat V_t\rangle_{\hat
t}, \ee where $\hat V_t\equiv \hat V_{t\tilde r}+\hat V_{t\tilde
\theta}$, cf. Eq. (\ref{dtmino}), and similarly for $\dot L_z$ and
$\dot K$.

The explicit results we obtain using this method are given in
section \ref{sec2}, Eqs. (\ref{edotavg}), (\ref{lzdotavg}), and
(\ref{kdotavg}).

\subsection{Averaging method using the explicit parameterization of
Newtonian orbits}
 To perform the time-averaging using this method, we define a parameter
$\xi$ via \be \tilde r=\frac{p}{1+e \cos \xi},\label{rxi0}\ee where
the parameter $\xi$ varies from $0$ to $2\pi$ as $\tilde r$ goes
through a complete cycle. Note that $\theta$ appears in Eqs.
(\ref{edotinst}) -- (\ref{kdotinst}) only in terms that are linear
in $Q$, so we can write $\theta$ in terms of $\xi$ using the
Newtonian relation \be x_3=r\cos\theta=r\sin\iota\sin(\xi+\xi_0).
\ee Here, $\xi_0$ is the angle between the direction of the
perihelion and the intersection of the orbital and equatorial plane.
Similarly, for the $\dot r \dot \theta$ terms in Eqs.
(\ref{lzdotinst}) and (\ref{Kdotinst}) we can use the Newtonian
relations $\dot r=e/\sqrt{p}\sin\xi$ and $\dot \xi= \sqrt{p}/r^2$.
From Eqs. (\ref{s2}) and (\ref{rxi}) it follows that
\begin{widetext}
 \be\frac{d\tilde t}{d\xi}=\frac{p^{3/2}}{(1+e \cos
 \xi)^2}\left\{ 1-\frac{Q}{8p^2}\left[-3+e^2-2e\cos \xi
 + 2 \cos^2 \iota
 (8-e^2+8e\cos \xi+e^2 \cos 2\xi )\right]\right\},
 \ee
 and from Eq. (\ref{ttildedef})
 \be
 \frac{dt}{d\tilde t}=\left\{1+\frac{Q}{2 p^2}\left
 (1+e \cos \xi\right)\left[2 \sin^2 \iota \sin^2 (\xi+\xi_0)
 -1\right]\right\}.
 \ee
Using these expressions, we compute the time-averaged fluxes from
\be \langle \dot E\rangle = \frac{\int^{2\pi}_0 d\xi {\;}\dot
E{\,}(dt/d\tilde t){\,}(d\tilde t/d\xi)}{\int^{2\pi}_0
d\xi{\,}(dt/d\tilde t){\,}(d\tilde t/d\xi)}\ee and obtain:

\noindent \bea \langle \dot E\rangle
=-\frac{32}{5}\frac{(1-e^2)^{3/2}}{p^5}&&\left[1+\frac{73}{24}e^2+\frac{37}{96}e^4
-\frac{S}{p^{3/2}}\left(\frac{73}{12}+\frac{823}{24}e^2+\frac{949}{32}e^4+\frac{491}{192}e^6\right)
\cos(\iota)\right.\nonumber\\
&&-\left.\frac{Q}{p^2}\left\{\frac{1}{2}+\frac{85}{32}e^2+\frac{349}{128}e^4+\frac{107}{384}e^6
+\left(\frac{11}{4}+\frac{273}{16}e^2+\frac{847}{64}e^4+\frac{179}{192}e^6\right)\cos
(2 \iota)\right\}\right.\nonumber\\
&&-\left.\frac{S^2}{p^2}\left\{\frac{13}{192}+\frac{247}{384}e^2+\frac{299}{512}e^4+\frac{39}{1024}e^6
-\left(\frac{1}{192}+\frac{19}{384}e^2+\frac{23}{512}e^4+\frac{3}{1024}e^6\right)\cos
(2 \iota)\right\}\right.\nonumber\\
&&-\left. \frac{ Q}{p^2}
e^2\left(\frac{869}{48}+\frac{1595}{96}e^2+\frac{121}{128}e^4\right)\cos
(2 \xi_0) \sin^2 \iota\right.\nonumber\\
&&+\left.\frac{S^2}{p^2}
e^2\left(\frac{1}{384}+\frac{5}{384}e^2+\frac{3}{2084}e^4\right)\cos
(2 \xi_0) \sin^2 \iota \right],\label{edotavgn}\eea

\noindent
\bea \langle \dot
L_z\rangle=-\frac{32}{5}\frac{(1-e^2)^{3/2}}{p^{7/2}}\cos
\iota&&\left[1+\frac{7}{8}e^2-\frac{S}{2
p^{3/2}\cos\iota}\left\{\frac{61}{24}+7e^2+\frac{271}{64}e^4+\left(\frac{61}{8}+\frac{91}{4}e^2+\frac{461}{64}e^4\right)
\cos(2\iota)\right\}\right.\nonumber\\
&&\left.-\frac{Q}{16p^2}\left\{-3-\frac{45}{4}e^2+\frac{19}{8}e^4
+\left(45+148e^2+\frac{331}{8}e^4\right)\cos
(2\iota)\right\}\right.\nonumber\\
&&\left.+\frac{S^2}{16p^2}\left\{1+3e^2+\frac{3}{8}e^4\right\}-\frac{Q}{p^2}e^2\cos
(2 \xi_0) \sin^2
\iota\left(\frac{201}{32}+\frac{51}{32}e^2\right)\right],\label{lzdotavgn}
\eea

\noindent
\bea \langle \dot
K\rangle=-\frac{64}{5}\frac{(1-e^2)^{3/2}}{p^{3}}&&\left[1+\frac{7}{8}e^2
-\frac{S}{2p^{3/2}}\left(\frac{97}{6}+37e^2+\frac{211}{16}e^4\right)\cos(\iota)\right.\nonumber\\
&&\left.-\frac{Q}{p^2}\left\{\frac{1}{2}+
\frac{55}{48}e^2+\frac{139}{192}e^4+\left(\frac{13}{4}+\frac{841}{96}e^2+\frac{449}{192}e^4\right)\cos
(2\iota)\right\}\right.\nonumber\\
&&\left.+\frac{S^2}{p^2}\left\{\frac{13}{192}+
\frac{13}{64}e^2+\frac{13}{512}e^4-\left(\frac{1}{192}+\frac{1}{64}e^2+\frac{1}{512}e^4\right)\cos
(2\iota)\right\}\right.\nonumber\\
&&\left.-\frac{Q}{p^2}
\left(\frac{391}{48}+\frac{37}{24}e^2\right)e^2\cos
(2\xi_0)\sin^2\iota\right].\label{kdotavgn} \eea \end{widetext} In
the adiabatic limit, the terms involving $\cos (2 \xi_0)$ can be
omitted because they average to zero. As explained by Ryan
\cite{ryan2}, the radiation reaction timescale for terms involving
$\xi_0$ is much longer than the precession timescale for most
orbits, so the terms involving $\xi_0$ will average away. This is
consistent with our results for the adiabatic infinite time-averaged
fluxes using the Mino time parameter. The Mino-time averaging method
was based on the assumption that the fundamental frequencies are
incommensurate and the motion fills up the whole torus, which is
equivalent to averaging over $\xi_0$.

\end{document}